\documentclass[12pt]{article}
\pdfoutput=1
\usepackage{savesym}
\usepackage{cite}
\usepackage{wasysym}
\usepackage{amscd}
\usepackage{graphicx}
\usepackage{pifont}
\usepackage{float}
\usepackage{subfig}
\usepackage[all]{xy}
\usepackage{multicol}
\usepackage{amsfonts}
\usepackage{picture}
\usepackage{amssymb}
\savesymbol{iint}
\savesymbol{iiint}
\usepackage{amsmath}
\usepackage{amsthm}

\newtheorem{prop}{Proposition}

\restoresymbol{TXF}{iint}
\restoresymbol{TXF}{iiint}
\usepackage{color}
\usepackage{clay}
\usepackage{hyperref}
\usepackage{slashed}
\hypersetup{colorlinks=true}
\hypersetup{linkcolor=black}
\hypersetup{citecolor=black}
\hypersetup{urlcolor=black}
\usepackage{setspace}
\usepackage{multirow}
\usepackage{wrapfig}
\usepackage{verbatim}
\numberwithin{equation}{section}

\begin{document}
\date{February, 2015}

\institution{Fellows}{\centerline{${}^{}$Society of Fellows, Harvard University, Cambridge, MA, USA}}
\title{Regge Trajectories in $\mathcal{N}=2$ Supersymmetric Yang-Mills Theory}

\authors{Clay C\'{o}rdova\let\thefootnote\relax\footnote{e-mail: {\tt cordova@physics.harvard.edu}}}

\abstract{We demonstrate that $\mathcal{N}=2$ supersymmetric non-Abelian gauge theories have towers of BPS particles obeying a Regge relation, $J \sim m^{2},$ between their angular momenta, $J,$ and their masses, $m$.   For $SU(N)$ Yang-Mills theories, we estimate the slope of these Regge trajectories using a non-relativistic quiver quantum mechanics model.  Along the way, we also prove various structure theorems for the quiver moduli spaces that appear in the calculation.}

\maketitle

\setcounter{tocdepth}{2}
\tableofcontents

\section{Introduction}

One of the most striking features of quantum chromodynamics is the existence of resonances of mesons and baryons with angular momenta, $J,$ and masses $m,$ lying on approximately linear Regge trajectories
\begin{equation}
J \sim\alpha ' \, m^{2}~. \label{r1}
\end{equation}

This Regge relationship between angular momentum and mass provides a conceptual link between non-Abelian gauge theory and string theory.  As first observed in \cite{Nambu:1969se, Nielsen, Susskind:1970xm}, a natural way to obtain \eqref{r1} is to consider a rotating rigid string whose ends move at the speed of light.  The mass and angular momentum then satisfy the Regge relation with
\begin{equation}
\alpha' = \frac{1}{2\pi T}~,
\end{equation}
where $T$ is the energy per unit length of the string.   This simple result is a starting point for a qualitative model of quark confinement where quark antiquark pairs are connected by the QCD string.  In the context of gauge-gravity duality \cite{Maldacena:1997re, Gubser:1998bc, Witten:1998qj}, the dual string picture of Yang-Mills is exact, and the Regge spectrum of operators in field theory may be deduced from spinning strings in Anti-de Sitter space \cite{Gubser:2002tv}.

In this work, we demonstrate that a relationship analogous to \eqref{r1} holds in $\mathcal{N}=2$ supersymmetric field theories.  Specifically we consider $SU(n_{c}+1)$ Yang-Mills theories with $n_{c}>1$.  These models are qualitatively distinct from their confining cousins with less supersymmetry.  The $\mathcal{N}=2$ theories have a moduli space of vacua and in a generic vacuum the low-energy physics is described by a Coulomb phase of free $u(1)$ vector multiplets.  Particles in these theories thus carry electric and magnetic charges $\gamma$ valued in an integral charge lattice $\Gamma$.  A complete solution to the spectrum of this theory is therefore an enumeration of the stable one-particle states in the Hilbert space for each electromagnetic charge $\gamma$. 

To state our result, we first fix a direction $\hat{\gamma}$ in the charge lattice $\Gamma$.  We examine particles with charges $\Lambda \hat{\gamma}$ where $\Lambda\gg1.$   We find that along this ray the stable particles of largest angular momentum obey the Regge relation
\begin{equation}
J \sim \alpha'(\hat{\gamma}) m^{2}~.\label{r2}
\end{equation}
In particular, these particles exist for $J$ arbitrarily large.  The slope function $\alpha'(\hat{\gamma})$ depends on the direction in the charge lattice $\hat{\gamma},$ and acquires its length scale from field expectation values $\langle \phi \rangle$ specifying the vacuum of the theory.   

In general in quantum field theory, the exact nature of the particle spectrum is difficult to determine.  In our case, we are able to obtain precise information by studying BPS particles, which preserve some of the underlying supersymmetry of the theory.   We model these BPS configurations as non-relativistic bound states described by quiver quantum mechanics.   In this picture the states saturating the Regge equation \eqref{r2}, are physically described by stable multi-centered configurations of dyons and monopoles and carry a large angular momentum in the induced electromagnetic fields.  We introduce these models in \S \ref{sec:toy} following the analysis in  \cite{Lee:1998nv, Douglas:2000ah, Douglas:2000qw, Denef:2002ru, Alim:2011kw}.

In \S \ref{sec:reg}, we investigate the bound state spectrum in non-Abelian gauge theory.  Our perspective is to view the BPS states in general $SU(n_{c}+1)$ gauge theory as composites of simpler states arising from various $SU(2)$ subgroups.   We demonstrate that the quiver moduli spaces describing classically stable states, are naturally decomposed into cells according to how the bound state in question is viewed as a multi-centered configuration of the distinct species of $SU(2)$ type dyons.  The states of largest angular momentum at fixed electromagnetic charges are then determined by the dimension of this moduli space.  Our main technical result is thus a sharp estimate for the dimension of the moduli space.

One of the interesting complexities of the spectrum of $\mathcal{N}=2$ field theories is the phenomenon of wall-crossing.  Depending on the expectation value $\langle \phi \rangle$ a given particle may or may not be stable.  The dynamics of the gauge theory also depends strongly on $\langle \phi \rangle$.  For large values of $\langle \phi \rangle,$ the non-Abelian Yang-Mills theory is broken to an Abelian theory at a high scale and is weakly coupled due to asymptotic freedom.  Our result \eqref{r2} holds in this region of parameter space.  By contrast, for small $\langle \phi \rangle$ the gauge theory is strongly coupled and there are only a finite number of stable BPS particles \cite{Lerche:2000uy, Fiol:2000pd, Alim:2011kw, Gaiotto:2012rg, Chuang:2013wt}. 

The weak coupling region of the $SU(n_{c}+1)$ gauge theory is divided into many chambers whose detailed spectra are different.  We are able to make progress in analyzing the particle content by working in a strict limit of parameters where there is a clear hierarchy of mass scales.  

To be specific, the conserved infrared charges are $n_{c}$ species of electric charge $e_{i},$ and $n_{c}$ species of magnetic charges $q_{i}:$  
\begin{equation}
\gamma= (e_{1}, e_{2}, \cdots, e_{n_{c}}, q_{1}, q_{2}, \cdots q_{n_{c}})\in \Gamma~.
\end{equation}
Each of the basic charges $e_{i}$ and $q_{j}$ has an associated complex mass scale set by the central charge $Z(\gamma)$.  The ratio of electric to magnetic mass scales is controlled by the effective fine structure constant which is parametrically small in the regime of interest
\begin{equation}
\frac{|Z(e_{i})|}{|Z(q_{i})|} \sim \frac{g^{2}}{4\pi} \ll 1~. \label{zas1}
\end{equation}
Meanwhile the relative mass scales for the distinct species of magnetic charge vary in the weak coupling region causing intricate jumping phenomena in the spectrum.  The structure simplifies in the limit where the central charges are phase ordered as\footnote{The ordering of the charges is not arbitrary, so this constraint is not trivial.  See \eqref{cartanpairing} and Figure \ref{fig:ymq} for a complete explanation.}
\begin{equation}
\arg(Z(q_{1}))> \arg(Z(q_{2}))> \cdots > \arg(Z(q_{n_{c}}))~, \label{phaseintro}
\end{equation}
and there is a parametric separation of scales
\begin{equation}
|Z(q_{i})| \gg |Z(q_{i+j})|~, \hspace{.5in} j>0~. \label{zas2}
\end{equation}
Alternative assumptions of hierarchies of the magnetic mass scales similarly yield simplifications in the BPS spectrum.

We work throughout in the limit where the inequalities \eqref{zas1} and \eqref{zas2} parametrically obeyed and neglect corrections to these approximations.  This is equivalent to working in a limit where there are parametrically controlled ratios of charges for the particles under consideration.  Thus, for all electric and magnetic charges we assume
\begin{equation}
\frac{e_{i}}{q_{i}} \ll \frac{ |Z(q_{i})|}{|Z(e_{i})|}\sim \frac{4\pi}{g^{2}}~. \label{cas1}
\end{equation}
And similarly we assume that for $j>0,$
\begin{equation}
\frac{q_{i}}{q_{i+j}}\gg \frac{|Z(q_{i+j})|}{|Z(q_{i})|}~. \label{cas2}
\end{equation}
The Regge trajectories, and other organizing features in the spectrum that we identify are correct when the charge restrictions \eqref{cas1} and \eqref{cas2} are obeyed.  When the charges become large enough to violate these bounds our results may receive  corrections.

With these preliminaries, we may now state our conclusions concerning the BPS spectrum in this regime of parameters.\footnote{We restrict ourselves to the case where all possible electric and magnetic charges are activated, so for each $i$, the pair $(e_{i},q_{i})\neq (0,0)$. Other restricted configurations of charges where $(e_{i},q_{i})=(0,0)$ for some $i$ may be understood inductively by decreasing $n_{c}$.} 
\begin{itemize}
\item In the weak coupling region, where the inequality \eqref{zas1} is obeyed, all stable BPS particles have either $q_{i}< 0 $ for all $i$, $q_{i}> 0$ for all $i$, or $q_{i}=0$ for all $i$.  The electric states, with $q_{i}=0$ for all $i,$ are perturbatively accessible and give rise to states with bounded angular momentum.  Meanwhile, the magnetically charged states have angular momentum which is unbounded.
\item If further we assume the hierarchies \eqref{phaseintro} and \eqref{zas2}, then we establish sufficient conditions for there to exist stable BPS states with a given collection of charges $\{e_{i}, q_{i}\}.$ Specifically:
\begin{itemize}
\item If all magnetic charges are positive, a sufficient condition for stable BPS states to exist is
\begin{equation}
\lceil e_{i}/q_{i}\rceil < \lfloor e_{i+1}/q_{i+1}\rfloor~, \hspace{.5in} i=1,\cdots, n_{c}-1~. \label{intexist1}
\end{equation}
\item If all magnetic charges are negative, a sufficient condition for stable BPS states to exist is
\begin{equation}
\lfloor e_{i}/|q_{i}|\rfloor > \lceil e_{i+1}/|q_{i+1}|\rceil~, \hspace{.5in} i=1,\cdots, n_{c}-1~.\label{intexist2}
\end{equation}
\end{itemize}
Where in the above, $\lceil x\rceil,$ and $\lfloor x\rfloor$ are the ceiling and floor of a real number $x$.

\item  For states with all positive magnetic charges satisfying \eqref{intexist1}, or all negative magnetic charges satisfying \eqref{intexist2}, we demonstrate the following lower bound on the angular momentum $J$ of BPS particles:\footnote{By the angular momentum of a BPS particle, we mean the largest angular momentum in the given BPS multiplet.  In particular, this implies that $J\geq 1/2$.}
\begin{equation}
J\geq \frac{1}{2}\left(\sum_{i=1}^{n_{c}-1}(e_{i+1}q_{i}-e_{i}q_{i+1})-\sum_{i=1}^{n_{c}}q_{i}^{2}\right)+\frac{1}{2}~. \label{intest}
\end{equation}
\end{itemize}

The Regge growth in the angular momentum may be deduced from these results.  Indeed, in any model where the BPS states can be described by quiver quantum mechanics, there is an a priori upper bound on the angular momentum of BPS particles in terms of a quadratic function of the charge $\gamma$. Thus, under scaling $\gamma \rightarrow \Lambda \gamma,$ the estimate \eqref{intest} implies that that $J$ scales as $\Lambda^{2}$.  On the other hand, the BPS bound implies that the masses $m$ of charged particles scale as $\Lambda,$ so parametrically \eqref{r2} is obeyed.

There are many questions left unanswered by the analysis in this work.  Of particular note however, is that our results do not shed any light on whether the Regge phenomenon we observe is connected to an underlying string interpretation.  We leave this, as well as a more thorough exploration of the sub-leading Regge trajectories and details of the spectrum as possibilities for future research.

\section{Toy Models}
\label{sec:toy}

In this section we develop a semiclassical intuition for the validity of the Regge relation \eqref{r2} in the BPS spectrum of $\mathcal{N}=2$ field theories.  Additionally, we introduce a non-relativistic quiver quantum mechanics toy model where simple calculations are feasible.  

Throughout this section, we take a low-energy point of view on the BPS spectrum of $\mathcal{N}=2$ field theories.  The BPS states are viewed as heavy dyons which interact through the long-range Coulomb and scalar interactions mediated by the massless vector multiplets.

\subsection{Semiclassical Intuition}
\label{secintuit}

Consider two particles with electric-magnetic charges given by $\gamma_{1}$ and $\gamma_{2}$ respectively.  Even if the individual particles do not carry angular momentum, the two particle configuration will carry angular momentum $J$ in the induced electromagnetic field given by the Dirac pairing between the two charges
\begin{equation}
J\sim \langle \gamma_{1}, \gamma_{2}\rangle~. \label{semiclassicalEM}
\end{equation}
Thus, if a bound state of these two particles exists, we anticipate that its angular momentum is semiclassically given by a quadratic function of the constituent charges.  Due to the linearity of electromagnetism, this phenomenon is general: given any multi-particle bound state the induced angular momentum in the electromagnetic fields is always a quadratic function of the charges of the particles.  To deduce Regge behavior, we will argue that the mass of such a bound state is a linear function of the charges.

For generic particles in quantum field theory, there is no a priori reason to expect a simple relationship between the mass of a particle and its electric and magnetic charges.  In this regard  $\mathcal{N}=2$ supersymmetric quantum field theories are different.  In these models, particles obey a bound relating mass and charge
\begin{equation}
m \geq | \sum_{i}(Z_{i}\gamma_{i})|~. \label{BPS}
\end{equation}
The complex parameters $Z_{i}$ are central charges and depend on the vacuum of the field theory as well as ultraviolet parameters like coupling constants and bare masses.

The bound \eqref{BPS} is saturated by BPS particles, which persevere some of the supersymmetry of the underlying quantum field theory.  They are the lightest states possible with given electromagnetic charge.  For our purposes, BPS particles are significant because, as is clear from \eqref{BPS}, the mass of such states is a linear function of the charges.  

We now combine these considerations with our semiclassical estimate of angular momentum.  Suppose that BPS particles exist with parametrically large charges of order $\Lambda$.  The mass of such states scales as $\Lambda$ while the angular momentum scales as $\Lambda^{2}$.  Thus we expect Regge behavior
\begin{equation} 
J \sim \alpha' \,m^{2}~.
\end{equation}
The slope $\alpha'$ above depends on the the central charges $Z_{i}$ which set the mass scale for the problem.  It furthermore depends on the charges $\gamma_{i}$ of the particles in question, but is homogenous under scaling $\gamma_{i}\rightarrow \Lambda \gamma_{i}$ .

From the previous analysis, we see that if such stable BPS particles exist it is natural to expect Regge scaling of the angular momentum and mass.  For this argument to be valid it is important that the bound states in question are semiclassical, with parametrically large charge and radius.  Meanwhile, stable non-BPS states, if they exist at all, must lie above the BPS bound \eqref{BPS}.  Assuming that their angular momentum is still dominated by the induced electromagnetic field we expect that these states lie below the leading Regge trajectory.

Despite the plausible argument given in this section, in a variety models, for instance $SU(2)$ gauge theory with arbitrary matter, the true microscopic description of BPS states allows for cancellations of angular momentum and the stable BPS states do not in fact form Regge trajectories \cite{Gaiotto:2009hg, Alim:2011ae}.  Thus, it is essential to give precise microscopic models for the BPS states where the issue of stability may be reliably addressed, and the angular momentum may be reliably computed.

\subsection{Quiver Quantum Mechanics Models}
\label{secquivertoy}

We now turn to concrete models realizing the semiclassical considerations of the previous section.  A simple quantitative picture may be developed using the non-relativistic quantum mechanics that governs the worldline theory of BPS particles.  In this section we briefly review these ideas in the context of a simple toy model.  For a more detailed introduction to these systems see \cite{Denef:2002ru, Alim:2011kw}.

We consider a system with two basic hypermultiplet BPS states, and take their charges to be $\gamma_{i}$, with electromagnetic product $k$
\begin{equation}
\langle \gamma_{1}, \gamma_{2}\rangle =k >0~.
\end{equation}

\begin{figure}[h!]
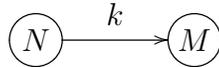

  \centering
\subfloat{
\xy  0;<1pt,0pt>:<0pt,-1pt>::
(-300,0) *+{N}*\cir<10pt>{} ="1",
(-240,0) *+{M}*\cir<10pt>{} ="2",
(-270, -10) *+{k} ="b",
\ar @{->} "1"; "2"
\endxy}
  \caption{The Kronecker quiver with dimension vector $(N,M)$ and $k$ arrows.  This model encodes the interactions of $N$ particles of charge $\gamma_{1}$ and $M$ particles of charge $\gamma_{2}$. }\label{fig:kronecker}
\end{figure}

These particles interact by long-range forces mediated by the exchange of scalars and vectors.  Due to supersymmetry the interaction Lagrangian for the multi-particle system is fixed to leading order in the particle velocities.  These interactions may be usefully encoded in a quiver quantum mechanics problem.  For the case in question the quiver is shown in Figure \ref{fig:kronecker}.

This diagram encodes a non-relativistic gauged quantum mechanics with four supercharges.  Each node is labelled by an integer and denotes a unitary gauge group of the indicated rank (i.e. $U(N)$ and $U(M)$ in the model shown in Figure \ref{fig:kronecker} ).  Meanwhile the arrow fields are chiral multiplets transforming in bifundamental representations.

The supersymmetric ground states of this quantum mechanics are BPS bound states of $N$ particles of charge $\gamma_{1}$ and $M$ particles of charge $\gamma_{2}$.  Thus, their total charge is given by
\begin{equation}
\gamma \equiv N\gamma_{1}+M\gamma_{2}~.
\end{equation}
In particular, stable bound states, if they exist, will have masses
\begin{equation}
m=|NZ_{1}+M Z_{2}|~, \label{masskron}
\end{equation}
where $Z_{i}$ are the central charges of this system.

Depending on the parameters $Z_{i}$ it may or may not be energetically favorable for the multi-particle system to form bound states.  When $Z_{1}/Z_{2}$ has argument in the lower half of the complex plane, the only stable states are the basic hypermultiplets with charges $\gamma_{1}$ and $\gamma_{2}$.  Meanwhile, when $Z_{1}/Z_{2}$ has argument in the upper half of the complex plane, an intricate spectrum of bound states exist.  These bound state spectra come in supermultiplets which are representations of the algebra $su(2)_{J}\times su(2)_{I}$, specifying the angular momentum ($J$) and $R$-symmetry ($I$) quantum numbers.  As representations, each supermultiplet takes the form
\begin{equation}
[(\mathbf{2},\mathbf{1})\oplus (\mathbf{1},\mathbf{2}) ] \otimes \mathbf{R}~, \label{centerom}
\end{equation}
where $\mathbf{R}$ is an irreducible representation of $su(2)_{J}\times su(2)_{I}.$  To determine the set of realized representations, we proceed as follows.

\begin{itemize}
\item Compute the classical moduli space $\mathcal{M}_{\gamma}$ parameterizing supersymmetric ground states of the quantum mechanics.  This moduli space is a K\"{a}hler quotient obtained by taking the vector space of constant configurations for the bifundamental chiral multiplet fields (indicated graphically by the arrows of the diagram), and quotienting by the gauge redundancy acting at each node of the diagram.\footnote{This is the Higgs branch approach to quantization.  For Coulomb branch approaches see \cite{Denef:2002ru, Manschot:2014fua}.}

One must also impose stability by removing certain loci from the space of field configurations.  We describe this procedure in detail in \S \ref{sec:stabdef}.

\item Quantize the moduli space $\mathcal{M}_{\gamma}$ by determining its cohomology.  As a vector space, the cohomology of $\mathcal{M}_{\gamma}$ is equal to a direct sum over all representations $\mathbf{R}$ occurring in \eqref{centerom} with the indicated charge $\gamma$.  Since $\mathcal{M}_{\gamma}$ is a K\"{a}hler manifold, the cohomology admits a Hodge decomposition with associated Hodge numbers $h^{p,q}$.  The integers $p$ and $q$ may in turn be interpreted in terms of the angular momentum and $R$-symmetry quantum numbers $J_{3}$ and $I_{3}$ as  
\begin{equation}
2J_{3}=p+q-\mathrm{dim}_{\mathbb{C}}(\mathcal{M}_{\gamma})~, \hspace{.5in}2I_{3}=p-q~. \label{qnform}
\end{equation}
In fact, for the special class of quiver models relevant in this paper, it is known that all cohomology classes have $p=q$ so that the $su(2)_{R}$ content of all representations $\mathbf{R}$ is trivial \cite{Chuang:2013wt, DelZotto:2014bga}.  Thus, the Hodge decomposition simply encodes the spin of BPS particles.

\end{itemize}
Form the above discussion, we see that determining the quantum numbers of BPS states is reduced to a cohomology problem for the moduli spaces $\mathcal{M}_{\gamma}.$  Mathematically, this is the study of quiver representation theory \cite{MR1315461}. 

In general, the complete spectrum of bound states of a quiver, including that of Figure \ref{fig:kronecker}, is complicated. Aspects of the spectrum of the toy model have been deduced by a variety of methods including equivariant cohomology \cite{2009arXiv0903.5442W, 2012arXiv1203.2740W}, wall-crossing formulas \cite{2008arXiv0811.2435K, 2009arXiv0903.0261R, 2009arXiv0909.5153G, Galakhov:2013oja}, and supersymmetric localization \cite{Cordova:2014oxa, Hori:2014tda, Ohta:2014ria, CS1, CS2}.  

However, to study Regge phenomena it suffices to extract the states of largest angular momentum given fixed charges.  This is a simpler task.  Indeed, from \eqref{qnform} it follows that the multiplet of maximal angular momentum is associated to the powers of the K\"{a}hler form.  The dimension of this representation is related the complex dimension of the moduli space $\mathcal{M}_{\gamma}$ as
\begin{equation}
\mathrm{dim}(\mathbf{R}_{max})=\mathrm{dim}_{\mathbb{C}}(\mathcal{M}_{\gamma})+1~. \label{jdim}
\end{equation}
Thus, in our study of Regge trajectories using quiver quantum mechanics, it suffices to determine the complex dimension of the classical moduli space $\mathcal{M}_{\gamma}$ for each choice of charge $\gamma$.

In the case of the toy model of Figure \ref{fig:kronecker} the complex dimension of the moduli space is easily computed by counting the number of degrees of freedom in the chiral multiplets (arrows), and quotienting by the (complexified) gauge redundancy.  We find\footnote{The offset is due to a central $GL(1,\mathbb {C})$ in the gauge group.} 
\begin{equation}
\mathrm{dim}_{\mathbb{C}}(\mathcal{M}_{\gamma})= k NM -N^{2}-M^{2}+1~. \label{dimform}
\end{equation}
Note that for $k<3$ the dimension above becomes negative for large $N$ and $M$.  This is a signal that no stable state with such charges exist. Conversely for $k\geq3$ the formula \eqref{dimform} correctly computes the dimension and is parametrically large as $N$ and $M$ become large.  

The large charge, and hence mass, of the states extracted for such moduli spaces suggests that the semiclassical argument for Regge behavior should be accurate.  Indeed, the highest spin BPS particle with the charge $\gamma$ has
\begin{equation}
J= \frac{1}{2}\left(k NM -N^{2}-M^{2}+1\right)~,
\end{equation} 
where the offset by $1/2$ is due to the tensor product in \eqref{centerom}.

Comparing to the formula for the mass \eqref{masskron} we find that for $N$ and $M$ large we obtain states with
\begin{equation}
J=\left[\frac{\left(k NM -N^{2}-M^{2}\right)}{2|NZ_{1}+M Z_{2}|^{2}}\right] m^{2}+\frac{1}{2}~.
\end{equation}
Note that the coefficient of $m^{2}$ in the above is a homogeneous function of the integers $M$ and $N$ specifying the charge $\gamma$.  Thus, if we fix the direction in the charge lattice and scale $(N,M)\rightarrow (\Lambda N, \Lambda M)$ for $\Lambda \gg1$ we obtain a Regge trajectory with slope function $\alpha'$ that depends only on the ratio $N/M$ specifying the direction in the charge lattice
\begin{equation}
\alpha' =\left[\frac{\left(k NM -N^{2}-M^{2}\right)}{2|NZ_{1}+M Z_{2}|^{2}}\right]  ~. \label{reggetoy}
\end{equation}

Above and beyond the interest in \eqref{reggetoy} as an exact result in this toy model, our calculation has broad implications for general $\mathcal{N}=2$ field theories.  Indeed, the toy model describes the bound states formed by spinless constituents whose electromagnetic charges form a two-dimensional lattice.  It may be embedded in a wide class of models, including $\mathcal{N}=2$ Yang-Mills theories with sufficiently large gauge group \cite{Cecotti:2012va, Galakhov:2013oja}.  In all such examples we therefore expect to find Regge trajectories in some region of moduli space. 

At the technical level, we can understand the occurrence of Regge trajectories in any theory where the BPS states may be described by a quiver model.   In such examples, the particles of largest angular momentum at fixed charge are determined through the dimension of the classical moduli space as in \eqref{jdim}.  If this moduli space is non-empty (a question which depends in detail on stability conditions governed by the central charges $Z_{i}$) then its dimension is given by a quadratic function of the charges obtained from the subtracting from the vector space of bifundamental fields, the dimension of the effective gauge group.  

In practice, the only complication in this logic is that the quiver moduli space may be subject to additional constraints arising from a superpotential which complicates the dimension calculation.  We confront this problem in the next section.

\section{Regge Trajectories in Super Yang-Mills}
\label{sec:reg}

In this section we turn to our main problem of interest: the spectrum of $SU(n_{c}+1)$ Yang-Mills in the weak coupling region of moduli space.  Like the toy problem of the previous section the bound state BPS spectrum of this theory is governed by a non-relativistic quiver quantum mechanics.  

The existence of a quiver model describing the BPS states implies that the angular momentum of a particle with charge $\gamma$ is bounded above by a quadratic function of $\gamma.$  Our aim is therefore to determine sufficient conditions for stable BPS states to exist, and to determine a lower bound on their angular momentum as a function of $\gamma$.

We begin in \S \ref{quivsunsec} by specifying this model in detail.  In particular, we introduce the electric and magnetic charges of BPS particles, and identify the weak coupling limit of parameter space in terms of central charges.  

In \S \ref{secsu2reps} we specialize to the case of $SU(2)$ and describe the BPS particles in that theory in the language of quiver representations.  

In \S \ref{su2gluesec}, we build on these results by viewing the general BPS state in  $SU(n_{c}+1)$ Super Yang-Mills, as a multi-centered configuration composed of the $SU(2)$ dyonic constituents.  To this end, we prove a number of results concerning the possible collections of $SU(2)$ dyons which may bind together into stable BPS particles.

Finally, in \S \ref{sec:regge}, we obtain our main results for the dimension of quiver moduli spaces, and establish an estimate for the Regge slope.

\subsection{A Quiver for $SU(n_{c}+1)$ Super Yang-Mills}
\label{quivsunsec}

On the Coulomb branch of moduli space, the gauge group $SU(n_{c}+1)$ is broken to $U(1)^{n_{c}}$ by the expectation value $\langle \phi \rangle$ of an adjoint scalar field.  The charge lattice $\Gamma$ of the theory is rank $2n_{c}$ and described as follows.   Let $\mathfrak{t}$ denote the Cartan subalgebra of $SU(n_{c}+1)$.  It is spanned by elements $\mathfrak{t}_{s}$ for $s=1, \cdots, n_{c}+1$ subject to a single constraint  
\begin{equation}
\mathfrak{t}=\left \{\beta_{1}\mathfrak{t}_{1}+\beta_{2}\mathfrak{t}_{2}+\cdots +\beta_{n_{c}+1}\mathfrak{t}_{n_{c}+1}| \sum_{s}\beta_{s}=0\right\}~.
\end{equation}
Similarly the dual space $\mathfrak{t}^{*}$ is spanned by elements $\mathfrak{t}^{*}_{s}$ subject to an analogous constraint.  The natural pairing between $\mathfrak{t}^{*}$ and $\mathfrak{t}$ is
\begin{equation}
\mathfrak{t}^{*}_{u}\cdot \mathfrak{t}_{v}=\delta_{u,v}~.
\end{equation}
Electric charges transform as weights of the Lie algebra and hence are valued in $\mathfrak{t}^{*},$ while magnetic charges are valued in $\mathfrak{t}$.   We express (electric, magnetic) charges as a pair valued in $(\mathfrak{t}^{*}, \mathfrak{t})$. 

The BPS spectrum of $SU(n_{c}+1)$ Yang-Mills theory may be viewed as non-relativistic bound states of $n_{c}$ species of monopoles and $n_{c}$ species of dyons.  Write $\gamma_{mi}$ for the charge of $i$-th monopole, $\gamma_{di}$ for the $i$-th dyon.  In the conventions introduced above these charges are
\begin{eqnarray}
\gamma_{mi} & = &(0,t_{i}-t_{i+1})~, \label{charges}  \\
\gamma_{di} & =& (t^{*}_{i}-t^{*}_{i+1},t_{i+1}-t_{i}) ~. \nonumber
\end{eqnarray}
We describe a general particle of charge $\gamma$ as a bound state by writing
\begin{equation}
\gamma=\sum_{i=1}^{n_{c}}\left(M_{i}\gamma_{m_{i}}+N_{i}\gamma_{di}\right)~,
\end{equation}
and viewing such a state as a composite of $M_{i}$ monopoles of type $i$ and $N_{i}$ dyons of type $i$.  

One may alternatively parameterize the particles in terms of more standard electric charges $e_{i}$ and magnetic charges $q_{j}$.  These are defined as
\begin{equation}
q_{j}\equiv M_{j}-N_{j}~, \hspace{.5in} e_{i}\equiv N_{i}~. \label{elecmagnet}
\end{equation} 
The charges vectors associated to this parametrization are respectively given by the monopole charge $\gamma_{mi}$ as well as the electric charge vectors $\gamma_{ei}\equiv (\gamma_{di}+\gamma_{mi})$.  Their pairings are 
\begin{equation}
\gamma_{ei}\cdot \gamma_{m_{j}}=C_{ij}~, \label{cartanpairing}
\end{equation}
where $C_{ij}$ denotes the Cartan matrix of the $SU(n_{c}+1)$ Lie algebra.  

In the following, we use the basis $\{M_{i}, N_{j}\}$ and $\{e_{i},q_{j}\}$ interchangeably.  We restrict our attention to bound states with general electromagnetic charges. Thus, we assume that for each $i,$ we have $(e_{i},q_{i})\neq (0,0).$  Special states, whose charges violate this assumption, can be understood inductively by examining smaller $n_{c}$.

\begin{figure}[h!]
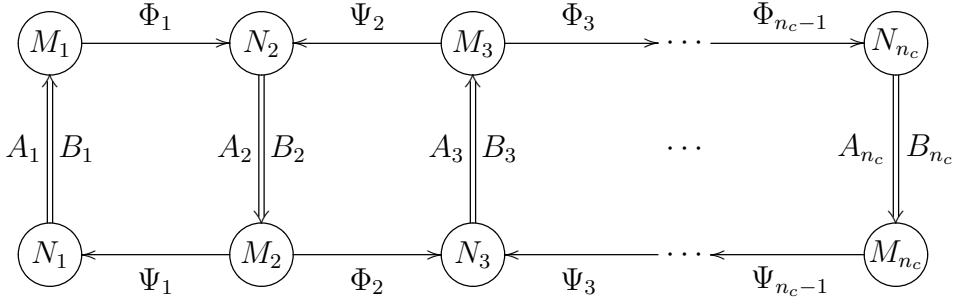

  \centering
\subfloat{
\xy  0;<1pt,0pt>:<0pt,-1pt>::
(0,80) *+{N_{1}}*\cir<12pt>{} ="1",
(0,0) *+{M_{1}}*\cir<12pt>{} ="2",
(80,80) *+{M_{2}}*\cir<12pt>{} ="3",
(80,0) *+{N_{2}}*\cir<12pt>{} ="4",
(160,80) *+{N_{3}}*\cir<12pt>{} ="5",
(160,0) *+{M_{3}}*\cir<12pt>{} ="6",
(240,80) *+{\cdots} ="7",
(240,0) *+{\cdots} ="8",
(320,80) *+{M_{n_{c}}}*\cir<12pt>{} ="9",
(320,0) *+{N_{n_{c}}}*\cir<12pt>{} ="10",
(240,40) *+{\cdots} ="11",
(-10, 40) *+{A_{1}} ="a",
(10, 40) *+{B_{1}} ="b",
(70, 40) *+{A_{2}} ="c",
(90, 40) *+{B_{2}} ="d",
(150, 40) *+{A_{3}} ="e",
(170, 40) *+{B_{3}} ="f",
(307, 40) *+{A_{n_{c}}} ="e",
(333, 40) *+{B_{n_{c}}} ="f",
(40, 90) *+{\Psi_{1}} ="g",
(120, 90) *+{\Phi_{2}} ="h",
(200, 90) *+{\Psi_{3}} ="i",
(280, 90) *+{\Psi_{n_{c}-1}} ="j",
(280, -10) *+{\Phi_{n_{c}-1}} ="k",
(200, -10) *+{\Phi_{3}} ="l",
(120, -10) *+{\Psi_{2}} ="m",
(40, -10) *+{\Phi_{1}} ="n",
\ar @{=>} "1"; "2"
\ar @{=>} "4"; "3"
\ar @{=>} "5"; "6"
\ar @{=>} "10"; "9"
\ar @{->} "2"; "4"
\ar @{->} "3"; "1"
\ar @{->} "3"; "5"
\ar @{->} "6"; "4"
\ar @{->} "6"; "8"
\ar @{->} "7"; "5"
\ar @{->} "9"; "7"
\ar @{->} "8"; "10"
\endxy}
  \caption{The quiver quantum mechanics problem for bound states of charge $\gamma$ in $SU(n_{c}+1)$ Yang-Mills.  The arrows are $A_{i}, B_{i}, \Phi_{i}, \Psi_{i}$ and label the bifundamental chiral multiplets in the model. }\label{fig:ymq}
\end{figure}

The interaction Lagrangian for our system is specified by a quiver quantum mechanics theory encoded in the Dirac pairing of the associated charges \cite{Fiol:2000pd, Alim:2011kw}.  It is illustrated in Figure \ref{fig:ymq}.
It is interesting that the fundamental non-Abelian electric degrees of freedom namely the W-bosons do not appear as elementary particles in the quiver model.  Instead, the vectors are produced as non-trivial bound states with enormous (negative) binding energy. The fact that W-bosons, which appear as fundamental fields in the Yang-Mills Lagrangian, do not play the role of elementary states is one of the most intriguing features of this approach to the spectrum.

Beyond the gauge interactions implied by the nodes and arrows, this quiver quantum mechanics has the additional complication of a non-trivial superpotential given by 
\begin{equation}
\mathcal{W}= \sum_{i=1}^{n_{c}-1}\left[\phantom{\int}\hspace{-.18in} \mathrm{Tr}\left(A_{i+1}\Phi_{i}A_{i}\Psi_{i}\right)-\mathrm{Tr}\left(B_{i+1}\Phi_{i}B_{i}\Psi_{i}\right)\right]~. \label{ncnfw}
\end{equation}
The moduli spaces $\mathcal{M}_{\gamma}$ of interest are thus subject to the constraint 
\begin{equation}
\frac{\partial \mathcal{W}}{\partial \chi}=0~, \label{ncnfw1}
\end{equation}
for all chiral multiplet fields $\chi$.

Finally, to full specify the model we must state the central charges of the fundamental monopolies and dyons.  These central charges may be extracted from the solution to the vector multiplet geometry \cite{Seiberg:1994rs, Klemm:1994qs}.  They take the form
\begin{equation}
Z_{mi}=\zeta_{i}~, \hspace{.5in}Z_{di}=-\zeta_{i}+i\varepsilon_{i}~. \label{light0}
\end{equation}
In the weak coupling region of moduli space $\zeta_{i}$ and $\varepsilon_{i}$ are approximately real and satisfy the inequality
\begin{equation}
|\varepsilon_{i}| \ll|\zeta_{j}|~, \hspace{.5in} \forall \, i,j~. \label{light}
\end{equation}

We may readily extract the physical meaning of these parameters and constrains.  From the BPS bound \eqref{BPS}, we see that the $i$-th fundamental monopoles and dyons have mass $|\zeta_{i}|$ plus small corrections of order $\varepsilon_{i}$.  Meanwhile, the parameters $\varepsilon_{i}$ control the masses of vector W-bosons associated to positive simple roots of the Lie algebra.  The constraint \eqref{light}, then implies that the states of pure electric charge, the W-bosons, are parametrically light compared to the magnetically charged monopoles and dyons.

This restriction on the central charges dovetails with the traditional description of the weak coupling region of moduli space.  There is a large vacuum expectation value $\langle \phi \rangle$ for the adjoint Higgs field, and the W-bosons acquire mass through the Higgs mechanism
\begin{equation}
m_{W}\sim g \langle \phi \rangle~.
\end{equation}
Meanwhile, the states of non-vanishing magnetic charge are viewed as non-perturbative semiclassical solitons with masses
\begin{equation}
m_{sol}\sim \frac{4\pi}{g}\langle \phi \rangle~.
\end{equation}
The ratio $m_{W}/m_{sol}$ is then of order the effective fine-structure constant which is parametrically small at weak coupling.\footnote{To be more precise, the expectation value $\langle \phi \rangle$ has $n_{c}$ eigenvalues and hence contains many scales.  We discuss constraints on the relative sizes of these scales in \S \ref{sec:stabdef}.}

In our analysis in the following we will work in an approximation where the electric charge correction to the central charge is treated as infinitesimal.  This approach is valid provided that the particles in question do not carry sufficiently large electric charge to yield meaningful corrections to $Z$.  Thus, we assume that for all $i,$
\begin{equation}
\frac{e_{i}}{q_{i}} \ll\frac{|\zeta_{i}|}{|\varepsilon_{i}|}\sim \frac{4\pi}{g^{2}}~.
\end{equation}
In the formal weak coupling limit, the right-hand-side of the above tends to infinity and our results become exact.  When the effective coupling is small but not infinitesimal, our results receive corrections for states with sufficiently large electric charges.

Having specified a quiver model, the problem of the BPS spectrum of $SU(n_{c}+1)$ super Yang-Mills is now reduced to the study of the moduli spaces $\mathcal{M}_{\gamma}$ for various choices of charge $\gamma$.  We restrict ourselves to the modest goal of determining the leading Regge trajectory of BPS states and hence aim to compute the dimension of the moduli spaces $\mathcal{M}_{\gamma}$.  This problem is complicated for two independent reasons.
\begin{itemize}
\item The superpotential equations \eqref{ncnfw}-\eqref{ncnfw1} imply that the moduli spaces $\mathcal{M}_{\gamma}$ are cut out by an intricate set of equations.  Hence computing the dimension of $\mathcal{M}_{\gamma}$ is not straightforward.  In \S \ref{supsolve} we address this issue by  solving the superpotential constraints.
\item Depending on the value of the central charge parameters $\zeta_{i}$ and $\varepsilon_{i},$ strata of the moduli spaces may appear and disappear and hence states may decay.  In fact, it is known \cite{Chuang:2013wt} that in the weak coupling region, defined by \eqref{light0}-\eqref{light} there are many distinct chambers, and thus many walls of marginal stability where states appear or disappear. We discuss this issue of stability in \S \ref{sec:stabdef}.
\end{itemize}

\subsection{Detailed Description of $SU(2)$ States}
\label{secsu2reps}

Our basic point of view on BPS states in general $SU(n_{c}+1)$ gauge theory is to view them as bound states of states familiar from the BPS spectrum of the $SU(2)$ theory.  Thus we begin with a more detailed description of these states.  The relevant quiver is a simplification of that shown in Figure \ref{fig:ymq}.
\begin{figure}[h!]
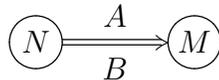

  \centering
\subfloat{
\xy  0;<1pt,0pt>:<0pt,-1pt>::
(-300,0) *+{N}*\cir<10pt>{} ="1",
(-240,0) *+{M}*\cir<10pt>{} ="2",
(-270, -10) *+{A} ="b",
(-270, 10) *+{B} ="b",
\ar @{=>} "1"; "2"
\endxy}
  \caption{The quiver for $SU(2)$ SYM with dimension vector $(N,M).$  This model Lagrangian encodes the interactions of $N$ dyons and $M$ monopoles. }\label{fig:kronecker1}
\end{figure}

Note that this quiver is a special case of the toy model explored in \S \ref{secquivertoy}.  However, in the case at hand, the number of bifundamental fields (arrows) is small and as a result the spectrum does not form Regge trajectories.  In fact the highest spin stable states are the vector bosons.  The remaining states are massive hypermultiplets carrying net magnetic charge $\pm 1$.  These states are familiar from a soliton analysis in field theory \cite{Sen:1994yi, Seiberg:1994rs ,Sethi:1995zm}, and may also be recovered by geometric methods from string theory \cite{Klemm:1996bj, Mikhailov:1998bx, Gaiotto:2009hg}.\footnote{For the connection between the geometric techniques and the quiver quantum mechanics studied here, see \cite{Alim:2011ae}.}

In the following, we present the $SU(2)$ spectrum as they appear from a quiver analysis.  Thus, we aim to describe the configuration of chiral multiplets, up to gauge equivalence, which specifies these moduli spaces.  

Mathematically the objects that we present in the following are \emph{representations} of the quiver in Figure \ref{fig:kronecker1}, and it is natural to describe them in that language \cite{MR1315461}.  Thus, the chiral fields (arrows) are viewed as linear maps between vector spaces supported at the nodes of the quiver.  Gauge transformations act as isomorphisms of these vectors spaces.  Finally the pair $(N,M)$ specifying the dimensions of the vector spaces are referred to as the \emph{dimension vector} of the given representation.  

Phrased in this language, the remainder of \S \ref{secsu2reps} is a complete description of the indecomposable representations of the quiver of Figure \ref{fig:kronecker1}.  All representations may be may be decomposed into a direct sum of these indecomposable building blocks.

\subsubsection{Dimension vector $(n,n+1)$}

The first possibility is a representation of dimension vector $(n,n+1)$.  These objects describe hypermultiplet dyons.  Their electric and magnetic charges are $(e,q)=(n,1).$
\begin{equation}
\begin{xy}  0;<1pt,0pt>:<0pt,-1pt>::
(0,0)*+{\mathbb{C}^{n}} ="1",
(60,0)*+{\mathbb{C}^{n+1}} ="2",
(30,9)*+{B},
(30,-9)*+{A},
"1", {\ar"2" <2.5pt>},
"1", {\ar"2" <-2.5pt>},
\end{xy}
\label{Drep}
\end{equation}

There is a unique indecomposable representation of this type for each $n\geq 0$.  We may characterize it by introducing a basis for the two vector spaces in question and specifying the behavior of the linear maps in this basis.  Thus, let $v_{1}, \cdots, v_{n}$ specify a basis at the source node and $w_{1}, \cdots, w_{n+1}$ denote a basis of vectors at the sink node.  The maps $A$ and $B$ are given by
\begin{equation}
A(v_{i})=w_{i}, \hspace{.5in}B(v_{i})=w_{i+1}.
\end{equation}
Alternatively, we may characterize this representation in a basis independent fashion by noting that the maps $A$ and $B$ have the property that there is no nontrivial subspace $\mathbb{C}^{k}\subset \mathbb{C}^{n}$ on which $A$ and $B$ agree.

We denote this representation in the following by $\mathcal{I}_{n}.$

\subsubsection{Dimension vector $(n,n-1)$}

The second type of indecomposable Kronecker representation has dimension vector $(n,n-1).$  These objects describe hypermultiplet dyons. Their electric and magnetic charges are $(e,q)=(n,-1).$
\begin{equation}
\begin{xy}  0;<1pt,0pt>:<0pt,-1pt>::
(0,0)*+{\mathbb{C}^{n}} ="1",
(60,0)*+{\mathbb{C}^{n-1}} ="2",
(30,9)*+{B},
(30,-9)*+{A},
"1", {\ar"2" <2.5pt>},
"1", {\ar"2" <-2.5pt>},
\end{xy}
\label{Erep}
\end{equation}
Up to isomorphism there is a unique indecomposable representation of this type for each $n\geq 1$.  

We again characterize the representation in a basis.  Let $v_{1}, \cdots, v_{n}$ specify a basis at the source node and $w_{1}, \cdots, w_{n-1}$ denote a basis of vectors at the sink node.  The maps $A$ and $B$ are given by
\begin{equation}
A(v_{i})= \begin{cases} w_{i} & i\leq n-1 \\ 0 & i=n\end{cases}, \hspace{.5in}B(v_{i})= \begin{cases} 0 & i=1 \\ w_{i-1} & i>1\end{cases}.
\end{equation}

We denote this representation in the following by $\mathcal{S}_{n}.$

\subsubsection{Dimension vector $(n,n)$}
\label{sec:w}

The final type of indecomposable Kronecker representation has dimension vector $(n,n).$ Their electric and magnetic charges are $(e,q)=(n,0).$ For $n=1$ they physically describe the vector W-boson.  For $n>1$ these states are only marginally stable and do not correspond to single particle states in the spectrum.  
\begin{equation}
\begin{xy}  0;<1pt,0pt>:<0pt,-1pt>::
(0,0)*+{\mathbb{C}^{n}} ="1",
(60,0)*+{\mathbb{C}^{n}} ="2",
(30,9)*+{B},
(30,-9)*+{A},
"1", {\ar"2" <2.5pt>},
"1", {\ar"2" <-2.5pt>},
\end{xy}
\label{Wrep}
\end{equation}
Up to isomorphism, these representations are labelled by an element $\lambda \in \mathbb{P}^{1}$.  The fact that the representation is non-rigid means that the associated particles carry spin in accordance with  \eqref{jdim}.

We again characterize these representations in a basis.  Let $v_{1}, \cdots, v_{n}$ specify a basis at the source node and $w_{1}, \cdots, w_{n}$ denote a basis of vectors at the sink node.  The map $B$ is then the identity matrix, while the map $A$ is a single Jordan block with eigenvalue $\lambda$
\begin{equation}
A=\left(\begin{array}{ccccc}\lambda & 1 & 0 & \cdots & 0 \\ 0 & \lambda & 1 & \cdots & 0 \\ \vdots & \vdots & \vdots & \ddots & \vdots \\ 0 & 0 & 0 & \cdots & 1 \\ 0 & 0 & 0 & \cdots & \lambda \end{array}\right)~, \hspace{.5in} B =\left(\begin{array}{ccccc}1& 0 & 0 & \cdots & 0 \\ 0 & 1 & 0 & \cdots & 0 \\ \vdots & \vdots & \vdots & \ddots & \vdots \\ 0 & 0 & 0 & \cdots & 0 \\ 0 & 0 & 0 & \cdots & 1 \end{array}\right)~.
\end{equation}

We denote this representation in the following by $\mathcal{V}_{n}.$

\subsection{$SU(n_{c}+1)$ States from Binding $SU(2)$ Dyons}
\label{su2gluesec}

We now address the BPS spectrum of $SU(n_{c}+1)$ governed by the general quiver representations appearing in Figure \ref{fig:ymq}.  Our strategy is to examine how such quiver representations decompose along the various $SU(2)$ subquivers visible in Figure \ref{fig:ymq}.  We then analyze the superpotential and stability constraints on how these constituents may bind together. 

Our first step is to ignore the maps $\Phi_{i}$ and $\Psi_{k}$.  When that is done, the quiver splits into $n_{c}$ disconnected two-node quivers.  Each of these two-node quivers is identical to that illustrated in Figure \ref{fig:kronecker1}, which governs the spectrum of $SU(2)$ SYM.   Therefore our detailed discussion of the $SU(2)$ quiver representations in \S \ref{secsu2reps} may be brought to bear on the general $SU(n_{c}+1)$ states.

Consider the $i$-th quiver shown below.
\begin{equation}
\begin{xy}  0;<1pt,0pt>:<0pt,-1pt>::
(0,0)*+{\mathbb{C}^{N_{i}}} ="1",
(60,0)*+{\mathbb{C}^{M_{i}}} ="2",
(30,9)*+{B_{i}},
(30,-9)*+{A_{i}},
"1", {\ar"2" <2.5pt>},
"1", {\ar"2" <-2.5pt>},
\end{xy}
\label{splitrep}
\end{equation}
Denote by $K^{i}$ the representation above.  

According to the analysis of the \S \ref{secsu2reps}, the representation $K^{i}$ may be split into a direct sum.  Thus, introduce natural numbers $n^{i}_{\alpha}, \ell^{i}_{\beta},$ and $p^{i}_{\gamma}$ indexing the representation appearing in the summation.  We have
\begin{equation}
K^{i}= \left(\bigoplus_{\alpha=1}^{s^{i}} \mathcal{I}_{n^{i}_{\alpha}}\right) \oplus \left(\bigoplus_{\beta=1}^{t^{i}} \mathcal{S}_{\ell^{i}_{\beta}}\right) \oplus \left(\bigoplus_{\gamma=1}^{u^{i}} \mathcal{V}_{p^{i}_{\gamma}}\right)~. \label{isum}
\end{equation}
Explicitly, this implies that the maps $A_{i}$ and $B_{i}$ are block diagonal, where each block appears as in \S \ref{secsu2reps}. 

Physically speaking, the decomposition \eqref{isum} means that the representation $K^{i}$ describes a multi-particle configuration of dyons with positive magnetic charge of type $i$, indexed by $n^{i}_{\alpha},$ dyons with negative magnetic charge of type $i$, indexed by $\ell^{i}_{\beta},$ and W-bosons of type $i,$ indexed by $p^{i}_{\gamma}$.  The maps $\Phi_{j}$ and $\Psi_{k}$ will bind together these constituents of $K^{i}$ for neighboring values of $i$. Thus, our physical picture of an $SU(n_{c}+1)$ BPS particle is a multi-centered configuration of dyonic states, where each center is a stable dyon or W-boson of various $SU(2)$ subgroups.

\subsubsection{Superpotential Constraints on Dyon Binding: Part I}
\label{supsolve}

Next, we reintroduce the maps $\Phi_{j}$ and $\Psi_{k},$ which providing the binding between the representations described above. Their properties are constrained by the superpotential \eqref{ncnfw}.  

Consider the portion of Figure \ref{fig:ymq} illustrated below.
\begin{equation}
\begin{xy}  0;<1pt,0pt>:<0pt,-1pt>::
(0,0)*+{\mathbb{C}^{N_{i}}} ="1",
(60,0)*+{\mathbb{C}^{M_{i}}} ="2",
(30,10)*+{B_{i}},
(30,-10)*+{A_{i}},
(90,-10)*+{\Phi_{i}},
(120,0)*+{\mathbb{C}^{N_{i+1}}} ="3",
(180,0)*+{\mathbb{C}^{M_{i+1}}} ="4",
(150,10)*+{B_{i+1}},
(150,-10)*+{A_{i+1}},
"1", {\ar"2" <2.5pt>},
"1", {\ar"2" <-2.5pt>},
"2", {\ar"3" <0pt>},
"3", {\ar"4" <2.5pt>},
"3", {\ar"4" <-2.5pt>},
\end{xy}~.
\label{dw1}
\end{equation}
The superpotential equation $\partial\mathcal{W}/\partial \Psi_{i}=0,$ yields the following constraint on the map $\Phi_{i}$
\begin{equation}
A_{i+1}\circ \Phi_{i}\circ A_{i}=B_{i+1}\circ \Phi_{i}\circ B_{i}~. \label{phiconst}
\end{equation}
Similarly, by varying $\mathcal{W}$ with respect to $\Phi_{i}$ we obtain the relation
\begin{equation}
A_{i}\circ \Psi_{i}\circ A_{i+1}=B_{i}\circ \Psi_{i}\circ B_{i+1}~, \label{psiconst}
\end{equation}
which constrains the following segment of Figure \ref{fig:ymq}
\begin{equation}
\begin{xy}  0;<1pt,0pt>:<0pt,-1pt>::
(0,0)*+{\mathbb{C}^{N_{i+1}}} ="1",
(60,0)*+{\mathbb{C}^{M_{i+1}}} ="2",
(30,10)*+{B_{i+1}},
(30,-10)*+{A_{i+1}},
(90,-10)*+{\Psi_{i}},
(120,0)*+{\mathbb{C}^{N_{i}}} ="3",
(180,0)*+{\mathbb{C}^{M_{i}}} ="4",
(150,10)*+{B_{i}},
(150,-10)*+{A_{i}},
"1", {\ar"2" <2.5pt>},
"1", {\ar"2" <-2.5pt>},
"2", {\ar"3" <0pt>},
"3", {\ar"4" <2.5pt>},
"3", {\ar"4" <-2.5pt>},
\end{xy}~.
\label{dw2}
\end{equation}

The equations \eqref{phiconst} and \eqref{psiconst} are the key constraints which we must solve in order to determine the dimension of the moduli space and hence the leading Regge trajectory. The remaining superpotential constraints arising from varying the maps $A_{i}$ and $B_{i}$ are studied in \S \ref{supsolveII}.

At the $i$-th quiver we have specified the behavior of the maps $A_{i}$ and $B_{i}$ through the decomposition \eqref{isum}, and the detailed description of the representations $\mathcal{I}_{n},$ $\mathcal{S}_{\ell},$ and $\mathcal{V}_{p}$.  Therefore, we may solve the constraints \eqref{phiconst} and \eqref{psiconst} on the maps $\Phi_{i}$ and $\Psi_{i}$ by simply evaluating on the preferred basis vectors of \S \ref{secsu2reps}.  We carry out this procedure in detail in Appendix \ref{glueme}.  The results are summarized below.
\begin{itemize}
\item Evaluated on each of the summands of \eqref{isum}, the binding maps $\Phi_{i}$ and $\Psi_{i+1}$ are semi-decreasing with respect to magnetic charge.  Specifically this means the following.
\begin{itemize}
\item The map $\Phi_{i}$ restricted to a summand $\mathcal{S}_{n^{i}_{\alpha}}$  (negative magnetic charge) can only have non-vanishing image inside representations $\mathcal{S}_{\ell^{i+1}_{\beta}}.$  The image of $\mathcal{S}_{n^{i}_{\alpha}}$ inside the representations $\mathcal{I}_{\ell^{i+1}_{\gamma}},$ (positive magnetic charge) and  $\mathcal{V}_{\ell^{i+1}_{\delta}}$ (vanishing magnetic charge) is necessarily zero.
\item The map $\Phi_{i}$ restricted to a summand $\mathcal{V}_{n^{i}_{\alpha}}$  (zero magnetic charge) can only have non-vanishing image inside representations $\mathcal{S}_{\ell^{i+1}_{\beta}},$ and $\mathcal{V}_{\ell^{i+1}_{\gamma}}$.  The image of $\mathcal{V}_{n^{i}_{\gamma}}$ inside the representations $\mathcal{I}_{\ell^{i+1}_{\gamma}},$ (positive magnetic charge) is necessarily zero.
\end{itemize}
Identical restrictions hold for the maps $\Psi_{j}.$
\item The number of complex parameters in a map $\Phi_{i}$ or $\Psi_{j}$ from $\mathcal{S}_{n}$ to $\mathcal{S}_{\ell}$ is $n-\ell$.  In particular, the map must vanish if $\ell \geq n$.  Moreover if the map is nonzero, then it is surjective.
\item The number of complex parameters in a map $\Phi_{i}$ or $\Psi_{j}$ from $\mathcal{I}_{n}$ to $\mathcal{I}_{\ell}$ is $\ell-n$.  In particular, the map must vanish if $n \geq \ell$.  Moreover if the map is nonzero, then it is injective.
\end{itemize} 

As we argue in the following subsection, these superpotential constraints dramatically constrain the possible stable BPS particles.

\subsubsection{Enforcing Stability}
\label{sec:stabdef}

In order to complete our task of evaluating the dimension of classical moduli space, we must at last confront the question of stability.  Consider a collection of dyons and W-bosons of various $SU(2)$ subgroups bound together by the gluing maps $\Phi_{i}$ and $\Psi_{k}.$ When does the corresponding quiver representation describe a stable single-particle state?

In general in the context quiver models of BPS states, there is a complete representation theoretic criterion to address this question \cite{Douglas:2000ah, 2006math.11510B}.  Given a quiver with specified ranks $D_{i}$ of gauge groups at the nodes and central charges $Z_{i}$ we say that the central charge of the associated quiver representation $R$ is 
\begin{equation}
Z(R)\equiv \sum_{i}D_{i}Z_{i}~.
\end{equation}
The condition of stability that we now impose, constrains the central charge $Z(R)$ compared to central charges of candidate decay channels of particles obtained by quantization of $R$. 

Specifically, we define a subrepresentation $S$ of $R$ to be a representation of the given quiver where the vector spaces at the nodes for $S$ are subspaces of those associated to $R,$ and the arrow maps of $S$ are defined by restriction from those of $R$.  The subrepresentation $S$ is potential decay mode of particles associated to $R$.  

Every representation $R$ has two trivial subrepresentations: the zero representation, which assigns the zero vector space to each node, and the identity subrepresentation, which is simply $R$ itself.  We say that $R$ is stable, as a representation, if for all non-trivial subrepresentations the following constraint on the phases of the central charges is obeyed
\begin{equation}
\arg(Z(S))<\arg(Z(R))~.\label{stab}
\end{equation}
It is the moduli space of stable quiver representations, as defined by \eqref{stab}, whose quantization yields the BPS spectrum.  Thus, our task now is to enforce stability, in addition to the superpotential constraints of the previous section, and to extract the dimension of the stable moduli space.

\begin{figure}[h!]
  \centering
  \includegraphics[width=0.8\textwidth]{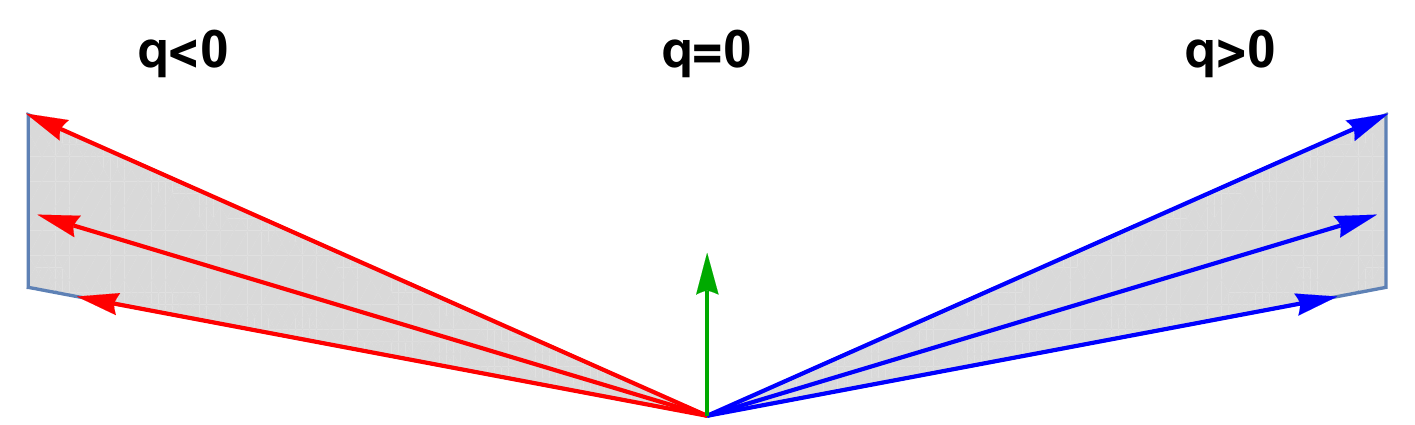}
  \caption{The central charge configuration at weak coupling.  The upper half of the complex $Z$-plane is shown and rays indicate possible central charges of representations of the $i$-th $SU(2)$ subquiver.  The representations are phase ordered according to their magnetic charge $q_{i}.$  Red rays have $q_{i}<0$ and have large argument.  Blue rays have $q_{i}>0$ and have small argument.  Electric states with $q_{i}=0$ are shown as green rays and have small absolute value.}
  \label{fig:zplane}
\end{figure}

To apply the above considerations to our analysis of BPS states in $SU(n_{c}+1)$ SYM, we must understand the geometry of the central charges \eqref{light0} in more detail.  The complex $Z$ plane is shown in Figure \ref{fig:zplane}.  Note in particular, that rays are phase ordered according to their magnetic charge.  That is, recalling the definition of the magnetic charge $q_{i}$ of \eqref{elecmagnet}, we have
\begin{equation}
q_{i}(R_{1})<q_{i}(R_{2}) \Longrightarrow \arg(Z(R_{1}))>\arg(Z(R_{2}))~.
\end{equation}

The phase ordering of representations according to their magnetic charge combines naturally with the superpotential constraints solved in \S \ref{supsolve} and yields severe restrictions on combinations of $SU(2)$ dyons which may be glued together to form stable bound states in $SU(n_{c}+1)$ SYM. 

 Indeed, consider the representations $K^{i}$ of equation \eqref{splitrep} and their associated direct sum decompositions \eqref{isum}.  From each $K^{i}$ we may extract the portion of the summand with $q_{i}<0,$ that is the direct sum over all representations of type $\mathcal{S}_{n^{i}_{\alpha}}$
\begin{equation}
W_{i}\equiv \left(\bigoplus_{\alpha=1}^{t^{i}} \mathcal{S}_{\ell^{i}_{\alpha}}\right)~.
 \end{equation}
 Clearly, $W_{i}$ is a subrepresentation of the $i$-th two-node subquiver.  Moreover, as a consequence of the superpotential constraints, the gluing maps $\Phi_{j}$ and $\Psi_{k}$ have the property that
 \begin{equation}
 \Phi_{i}\left(\mathrm{Sink}(W_{i})\right)\subset \mathrm{Source}(W_{i+1})~,\hspace{.5in}\mathrm{and} \hspace{.5in} \Psi_{i}\left(\mathrm{Sink}(W_{i+1})\right)\subset \mathrm{Source}(W_{i})~.
 \end{equation}
 We conclude that the union over $i$ of the $W_{i}$ forms a subrepresentation of the full candidate quiver representation of Figure \ref{fig:ymq}.  Hence we must impose the stability condition \eqref{stab}.  However, the $W_{i}$ consist only of representations with negative magnetic charge for each of the $i$ species.  If any $K^{i}$ contains a component with $q_{i}\geq 0$ then the subrepresentation we have just constructed will destabilize the moduli space.  Therefore, the only consistant possibility is that subrepresentation in question is trivial: either it is the entire object illustrated in Figure \ref{fig:ymq}, or it is the empty subrepresentation with vanishing vector spaces.
 
 In the case where the subrepresentation constructed above is empty, we proceed analogously using the components of $K^{i}$ with vanishing magnetic charge and construct another subrepresentation which may destabilize our configuration.  
 
 In this way we conclude that stable quiver representations of Figure \ref{fig:ymq} come in three broad types.
 \begin{enumerate}
 \item \textbf{Positive Magnetic Charge}
 
 Each representation $K^{i}$ has components containing only $q_{i}>0$
 \begin{equation}
 K^{i}= \bigoplus_{\alpha=1}^{s^{i}} \mathcal{I}_{\ell^{i}_{\alpha}} ~. \label{posrep}
 \end{equation}
   \item \textbf{Vanishing Magnetic Charge}
  
Each representation $K^{i}$ has components containing only $q_{i}=0$ 
\begin{equation}
 K^{i}= \bigoplus_{\gamma=1}^{u^{i}} \mathcal{V}_{\ell^{i}_{\gamma}}~.
 \end{equation} 
 
 \item \textbf{Negative Magnetic Charge}
 
 Each representation $K^{i}$ has components containing only $q_{i}<0$
 \begin{equation}
 K^{i}= \bigoplus_{\beta=1}^{t^{i}} \mathcal{S}_{\ell^{i}_{\beta}}~. \label{negrep}
 \end{equation}
 \end{enumerate}
 In particular, BPS states whose magnetic charges violate these constraints do not exist in the region of parameter space as defined by \eqref{light}.
 
 In the weak coupling region of interest, electric states (with all $q_{i}$ vanishing) are parametrically light compared to particles with nontrivial magnetic charges.  In particular, the electric objects are accessible via a perturbative Lagrangian analysis. Their spectrum consists of an $SU(n_{c}+1)$ adjoint of vector multiplets.  The multiplets associated to roots of the algebra are W-bosons and acquire mass through the Higgs mechanism, while the vectors associated to the Cartan subalgebra are massless on the Coulomb branch. As demonstrated in \cite{Alim:2011kw, Cecotti:2012va}, these results may be recovered using the more abstract quiver techniques employed here. 

We deduce in particular that the class of electric representations yields states with bounded spin.  Meanwhile, as we will subsequently demonstrate, the representations of total positive and negative electric charge yield towers of particles with unbounded spin.
 
We may further constrain the bound states of total positive and negative magnetic charge via a refined stability analysis.  To do so, we must be more specific about the values of the central charges.  So far we have focused on any of the two-node $SU(2)$ quivers and demanded that they are in a weak coupling regime of their central charges illustrated in Figure \ref{fig:zplane}.  However, we have made no specification of the relative properties of the central charges for the different species of dyons and monopoles.  Thus, let us recall our parametrization of the central charges 
\begin{equation}
Z_{mi}=-\zeta_{i}+i\varepsilon_{i}~, \hspace{.5in}Z_{di}=\zeta_{i}+i\varepsilon_{i}~. \label{light1}
\end{equation}
We now make a specification on the complex parameters $\zeta_{i}$ which specify the masses of magnetic charges.  Specifically, we assume that they are phase ordered
\begin{equation}
\arg(\zeta_{1})>\arg(\zeta_{2})> \cdots > \arg(\zeta_{n_{c}})~, \label{arghier}
\end{equation}
and moreover we work in a limit of parameters where there is a hierarchy of mass scales 
\begin{equation}
|\zeta_{1}|\gg|\zeta_{2}|\gg\cdots \gg|\zeta_{n_{c}}|~.\label{hier}
\end{equation}
Alternative orderings of the scales and phases provide similar simplifications in the spectrum.

The utility of the above restriction is that, in assessing stability using the definition \eqref{stab}, the contribution to the central charge is dominated first by the magnetic charge $q_{1},$ then by the magnetic charge $q_{2}$, and so on.  However, this line of logic may be violated if the magnetic charges become large enough to violate the hierarchy \eqref{hier}.  Thus, our analysis in the following will be restricted to charges obeying the constraint
\begin{equation}
\frac{q_{j}}{q_{i}}\ll\frac{|\zeta_{i}|}{|\zeta_{j}|}~, \hspace{.5in} j>i~.
\end{equation}
In the formal limit where the ratios on the right-hand-side of the above tend to infinity, our approximation becomes exact.\footnote{This approximation is similar to that used in \cite{Chuang:2013wt, Cordova:2013bza} to study framed BPS states.}

\subsubsection{Superpotential Constraints on Dyon Binding: Part II}
\label{supsolveII}

We now complete our analysis of the superpotential constraints by examining the equations derived from varying $\mathcal{W}$ of \eqref{ncnfw} with respect to $A_{i}$ and $B_{i}$.

We obtain the following constraints
\begin{eqnarray}
\Psi_{i}\circ A_{i+1}\circ \Phi_{i}+\Phi_{i-1}\circ A_{i-1}\circ \Psi_{i-1} & = & 0~,  \label{finalws} \\
\Psi_{i}\circ B_{i+1}\circ \Phi_{i}+\Phi_{i-1}\circ B_{i-1}\circ \Psi_{i-1} & = & 0~, \nonumber 
\end{eqnarray}
where in the above, we use the convention that maps appearing with subscripts outside the allowed range implied by Figure \ref{fig:ymq} are defined to be zero.

We claim that the equations \eqref{finalws}, combined with stability constraints \eqref{arghier}-\eqref{hier} imply the following simplification:
\begin{itemize}
\item In all stable representations of total positive magnetic charge, or total negative magnetic charge, the maps $\Psi_{i}$ vanish for all $i$.\footnote{The symmetry between $\Phi$ and $\Psi$ is broken by the choice of central charges \eqref{arghier}-\eqref{hier}.}
\end{itemize}

It is simple to demonstrate this claim.  Begin with the representation $K^{1}$ and consider its image under the map $\Phi_{1}$.    As a consequence of the detailed form of the representations described in \S \ref{secsu2reps}, the linear maps $A_{2}, B_{2}$ have the property that the direct sum of their image spans the entire vector space supported at their target node in the quiver.    From the equations \eqref{finalws}, we then deduce that $\Psi_{1}$ annihilates the image of $\Phi_{1}$ inside $K^{2}$.

Proceeding inductively, we similarly demonstrate that $\Psi_{i+1}$ annihilates the $i$th sequential image $\Phi_{i}\circ \Phi_{i-1}\cdots \circ \Phi_{1}(K^{1}).$  Thus, we construct a subrepresentation $S$  
\begin{equation}
S=K^{1}\rightarrow \Phi_{1}(K^{1})\rightarrow \Phi_{2}\circ \Phi_{1}(K^{1})\rightarrow \cdots \rightarrow \Phi_{n_{c}-1}\circ \cdots \circ \Phi_{2}\circ \Phi_{1}(K^{1})~. \label{nested}
\end{equation}
By construction, $S$ contains all of the representation $K^{1}$.  It follows that if we denote by $R$ the entire quiver representation, then we can find non-negative integers $\rho_{i}$ such that the central charges of $R$ and $S$ obey
\begin{equation}
Z(R)=Z(S)+\sgn(q)\sum_{i=2}^{n_{c}}\rho_{i} \zeta_{i}+\mathrm{electric \ terms}~, \label{finaldestab}
\end{equation}
where in the above $\sgn(q)=1$ on the representations of total positive magnetic charge and $\sgn(q)=-1$ on the representations of total negative magnetic charge.  But now from \eqref{arghier}-\eqref{hier}, we deduce that if $S$ is not equal to $R$, that is if the $\rho_{i}$ are non-zero, then the stability condition \eqref{stab} is violated.  We conclude that $S=R$ and hence all maps $\Psi_{i}$ are in fact zero.

Thus, the general quiver representation illustrated in Figure \ref{fig:ymq} is reduced to the representation illustrated below.
\begin{equation}
\begin{xy}  0;<1pt,0pt>:<0pt,-1pt>::
(0,0)*+{\mathbb{C}^{N_{1}}} ="1",
(60,0)*+{\mathbb{C}^{M_{1}}} ="2",
(30,10)*+{B_{1}},
(30,-10)*+{A_{1}},
(90,-10)*+{\Phi_{1}},
(120,0)*+{\mathbb{C}^{N_{2}}} ="3",
(180,0)*+{\mathbb{C}^{M_{2}}} ="4",
(150,10)*+{B_{2}},
(150,-10)*+{A_{2}},
(210,-10)*+{\Phi_{2}},
(270,-10)*+{\Phi_{n_{c}-1}},
(240,0)*+{\cdots} ="5",
(300,0)*+{\mathbb{C}^{N_{n_{c}}}} ="6",
(360,0)*+{\mathbb{C}^{M_{n_{c}}}} ="7",
(330,10)*+{B_{n_{c}}},
(330,-10)*+{A_{n_{c}}},
"1", {\ar"2" <2.5pt>},
"1", {\ar"2" <-2.5pt>},
"2", {\ar"3" <0pt>},
"3", {\ar"4" <2.5pt>},
"3", {\ar"4" <-2.5pt>},
"4", {\ar"5" <0pt>},
"5", {\ar"6" <0pt>},
"6", {\ar"7" <-2.5pt>},
"6", {\ar"7" <2.5pt>},
\end{xy}
\label{Full}
\end{equation}
Where in the above, the maps $\Phi_{i}$ are further restricted by the analysis of \S \ref{supsolve}.   

\subsection{Dimension Formulas for Quiver Moduli}
\label{sec:regge}

In this section we conclude our analysis by providing a cell decomposition of the moduli spaces of magnetically charged states.  Each cell corresponds to a way of viewing the BPS state as a multi-centered configuration of $SU(2)$ dyons.  We compute the dimension of these cells and provide a complete criterion to determine which cells lie in the stable moduli space.  

In \S \ref{secsmallmag} we investigate the resulting dimension formulas for states with small magnetic charges, and determine exact formulas for angular momenta.  In particular, we see that such states already have unbounded angular momentum.

Finally, in \S \ref{secsemi} we use our dimension formulas to obtain a simple lower bound on the Regge slope.

\subsubsection{Bound States of Dyons with Positive Magnetic Charge}
\label{secqposreg}

We consider first the class of representations with positive magnetic charge \eqref{posrep}.  We rewrite the representation $K^{i}$ of $i$-th two-node quiver.  Inserting the magnetic charge $q_{i}>0,$ and the electric charge $e_{i},$ this takes the form
\begin{equation}
\begin{xy}  0;<1pt,0pt>:<0pt,-1pt>::
(0,0)*+{\mathbb{C}^{e_{i}}} ="1",
(60,0)*+{\mathbb{C}^{e_{i}+q_{i}}} ="2",
(30,9)*+{B_{i}},
(30,-9)*+{A_{i}},
"1", {\ar"2" <2.5pt>},
"1", {\ar"2" <-2.5pt>},
\end{xy}~.
\label{splitrep3}
\end{equation}
Now enforce the direct sum decomposition \eqref{posrep} of $K^{i}$.  Using the fact that each representation $\mathcal{I}_{n}$ has dimension vector $(n,n+1)$, we see that the direct sum decomposition of $K^{i}$ must involve exactly $q_{i}$ summands.  

Said differently, there is a partition of the electric charge $e_{i}$ into $q_{i}$ non-negative integral parts.  We indicate the partition by its multiplicities.  Thus, define $\lambda_{i}(s)$ as the multiplicity with which the integer $s$ occurs
\begin{equation}
\sum_{s=0}^{e_{i}} \lambda_{i}(s) =q_{i}~, \hspace{.5in} \sum_{s=0}^{e_{i}}s \lambda_{i}(s) =e_{i}~. \label{partdata}
\end{equation}
The partition $\lambda_{i}$ specifies the direct sum decomposition of $K^{i}$ as
\begin{equation}
K^{i}=\left(\bigoplus_{\alpha=1}^{\lambda_{i}(0)}\mathcal{I}_{0}\right) \oplus \left(\bigoplus_{\alpha=1}^{\lambda_{i}(1)}\mathcal{I}_{1}\right) \oplus \cdots \oplus \left(\bigoplus_{\alpha=1}^{\lambda_{i}(e_{i})}\mathcal{I}_{e_{i}}\right) ~. \label{bigsplit}
\end{equation}

In other words, fixing the magnetic charge $q_{i}$ and a partition $\lambda_{i}$ specifies a $q_{i}$-centered configuration of dyons in the $i$-th $SU(2)$ subquiver.  

An important property of \eqref{bigsplit} is its automorphism group: the group of complexified gauge transformations at the nodes which stabilize the direct sum decomposition in question.  This is given by 
\begin{equation}
\mathrm{Aut}(K^{i})=\prod_{s}Gl(\lambda_{i}(s),\mathbb{C})~. \label{autogroup}
\end{equation}
It has complex dimension 
\begin{equation}
\mathrm{dim}\left(\mathrm{Aut}(K^{i})\right) =\sum_{s=0}^{e_{i}}\lambda_{i}(s)^{2}~.
\end{equation}

Now introduce the maps $\Phi_{j}$ connecting the distinct $K^{i}$.  The $\Phi_{j}$ satisfying the required superpotential constraints are constructed explicitly in Appendix \ref{glueme}.  In particular they have a known number of parameters as discussed in \S \ref{supsolve}.  Thus, we may now compute the dimension of the moduli space of representations \eqref{Full} where each $K^{i}$ has decomposition specified by the partition $\lambda_{i}.$ To do so, we simply sum over the parameters in the $\Phi_{i}$ and quotient by the automorphism group \eqref{autogroup}.  Denoting by $ \Delta_{q>0}\{\lambda_{i}\}$ the resulting dimension we find
\begin{equation}
\Delta_{q>0}(\left\{\lambda_{1}, \cdots \lambda_{n_{c}}\right\})= \sum_{i=1}^{n_{c}-1}\sum_{r=0}^{e_{i+1}}\sum_{s=0}^{r}(r-s)\lambda_{i+1}(r)\lambda_{i}(s) -\sum_{i=1}^{n_{c}}\sum_{s=0}^{e_{i}}\lambda_{i}(s)^{2}+1~, \label{qposdim1}
\end{equation}
where in the above, the offset by $1$ is due to an overall central $GL(1,\mathbb{C})$ in the automorphism group \eqref{autogroup}.

To proceed further, we must clarify the geometric meaning of the partitions $\lambda_{i}$ introduced above.  Let $\gamma^{+}$ denote the total charge as specified by the collection $\{e_{i},q_{i}\}$ (the superscript $+$ denotes that this is a state of total positive magnetic charge).  The moduli space of total charge $\gamma^{+}$ has cells described by the $n_{c}$ partitions $\lambda_{i}$.  The cells are glued together to form the total moduli space.  Each cell has dimension given by the formula \eqref{qposdim1}.  Strictly speaking, this is correct only if $\Delta_{q>0}$ is positive.   If the dimension computed by $\Delta_{q>0}$ is negative, then the automorphism group \eqref{autogroup} does not act freely, and the dimension of the cell is zero.  

The physical meaning of the decomposition of the moduli space into cells labelled by partitions is that a bound state of total charge $\gamma^{+}$ may be described as a multi-centered configuration of $SU(2)$ dyons in a variety of ways each determined by the given $n_{c}$-tuple of partitions.   

\begin{figure}[h!]
  \centering
  \subfloat[$\{\lambda_{i}\}\notin \mathcal{C}^{+}$]{\label{fig:ncon}\includegraphics[width=0.3\textwidth]{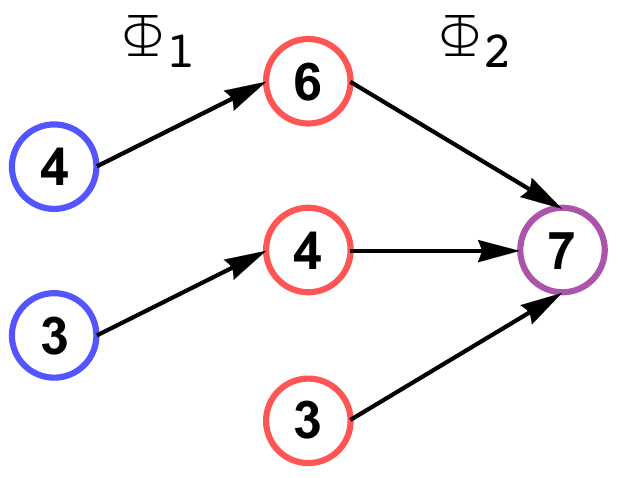}}
  \hspace{1in}
  \subfloat[$\{\lambda_{i}\}\in \mathcal{C}^{+}$]{\label{fig:con}\includegraphics[width=0.3\textwidth]{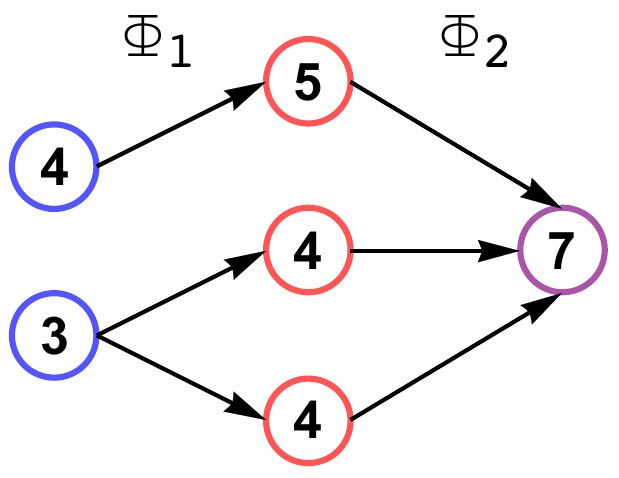}}
  \caption{An example of the connectivity properties on representations implied by stability for the case $n_{c}=3$, with charges $(e_{1},q_{1})=(7,2),$ $(e_{2},q_{2})=(13,3),$ and $(e_{3},q_{3})=(7,1).$  The elements of the partition of $e_{1}$ are blue, the elements of the partition of $e_{2}$ are red, and the elements of the partition of $e_{3}$ are purple.  The maps $\Phi_{1}$ and $\Phi_{2}$ are indicated by arrows.  In $(a),$ the collection of partitions is unstable because the element $3$ of the partition of $e_{2}$ is not in the image of any element of the partition of $e_{1}$.  In $(b),$ a tuple of partitions with the same charges which lies in $\mathcal{C}^{+}$.}
  \label{fig:connected}
\end{figure}

Not all $n_{c}$-tuples of partitions give rise to cells in the \emph{stable} moduli space.  Indeed, according to the analysis above equation \eqref{nested}, we must assume that the sequential images of $K^{1}$ via $\Phi_{i}$ generate the entire representation.   This means that for each element $n$ in the partition of $e_{i+1},$ there exists an element $m$ in the partition of $e_{i}$ with $n>m.$  We denote the set of partitions obeying this constraint as $\mathcal{C}^{+}.$    Any $n_{c}$-tuple of partitions which gives rise to a cell in the stable moduli space must lie in $\mathcal{C}^{+}$.  Representations whose associated partitions lie in $\mathcal{C}^{+}$ have strong connectivity properties as illustrated in Figure \ref{fig:connected}.

While this connectivity constraint is necessary, it is not in fact sufficient to ensure that a given $n_{c}$-tuple of partitions yields a cell in the stable moduli space.  Those tuples that must additionally be removed are such that the image of a dyon in say $K^{i}$ does not bind (via the map $\Phi_{i}$) to a sufficiently large fraction of the dyons in $K^{i+1}$.  To be specific, we may consider potentially destabilizing subrepresentations given by taking a collection of any $\ell_{1}$ of the dyon representations appearing in $K^{1}$, with $1\leq \ell_{1} < q_{1},$ and examining their sequential images under the maps $\Phi_{i}$ as in \eqref{nested}.  Denote the initial collection (a subset of $K^{1}$) by $\mathcal{I}_{n_{1}}, \cdots \mathcal{I}_{n_{\ell_{1}}}.$  The $n_{\alpha}$ are parts of the partition of $e_{1}$ (not to be confused with the multiplicities of these parts as given by $\lambda_{1}$). 

 Each $\mathcal{I}_{n_{\alpha}}$ above maps via $\Phi_{1}$ to those dyonic representations in $K^{2}$ which have larger electric charge.  Let $\ell_{2}$ denote the total number of dyonic subrepresenations of $K^{2}$ which lie in the image of $\Phi_{1}$ restricted to the subset $\mathcal{I}_{n_{1}}, \cdots \mathcal{I}_{n_{\ell_{1}}}.$  In terms of the partition $\lambda_{2}$ of $e_{2}$ the quantity $\ell_{2}$ may be defined as follows:  
\begin{equation}
\ell_{2}=\#\mathrm{elements}\left(\bigcup_{\alpha=1}^{\ell_{1}}\{s \in \mathrm{partition  \ of}  \ e_{2}|  \ s> n_{\alpha} \}\right)~.
\end{equation}
Similarly we define $\ell_{i+1}$ by considering the number of dyon subrepresentations of $K^{i+1}$ which lie in the image of $\Phi_{i}\circ \Phi_{i-1}\cdots \circ \Phi_{1}$ restricted to our initial collection of $\ell_{1}$ summands of $K^{1}$. Each of the $\ell_{i}$ are readily defined from the data of the tuple of partitions $\{\lambda_{1}, \cdots \lambda_{n_{c}}\}$ as with $\ell_{2}$ above.

As a simple example with $n_{c}=3,$ suppose that $(e_{1},q_{1})=(8,3),$ $(e_{2}, q_{2})=(16,5),$ and $(e_{3}, q_{3})=(26,2)$ with associated partitions $\{1,2,5\},$ $\{1,1,3,4,7\},$ and $\{1,25\}$.  If the initial $\ell_{1}=2$ with the chosen parts given by $n_{1}=1$, and $n_{2}=2$, then $\ell_{2}=3,$ and $\ell_{3}=1$. 

The sequential images of the original collection of $\ell_{1}$ dyonic subrepresentation of $K^{1}$ give rise to a subrepresentation $S$ as in \eqref{nested} which may potentially violate the stability constraint \eqref{stab}.  To ensure that it does not, we must demand  
\begin{equation}
\frac{q_{2}}{q_{1}}\leq \frac{\ell_{2}}{\ell_{1}} ~.\label{lineq1}
\end{equation}
If the inequality \eqref{lineq1} is strict, for all possible choices of $\ell_{1}$ initial components of $K^{1},$ then the representation is stable and the partition contributes to the moduli space $\mathcal{M}_{\gamma^{+}}.$  If the inequality \eqref{lineq1} is an equality, then we must demand further that 
\begin{equation}
\frac{q_{3}}{q_{1}}\leq \frac{\ell_{3}}{\ell_{1}}~. \label{lineq2}
\end{equation}
Again if the inequality is strict, then the representation is stable and the partition contributes to the moduli space.  In the case where it is saturated we must examine the next ratio $\ell_{4}/\ell_{1}$ and so on.  

To state the complete stability requirement on the $n_{c}$-tuple of partitions $\{\lambda_{i}\}$, we introduce  subsets $\upsilon^{+}_{i}$ of partitions defined as
\begin{eqnarray}
\upsilon^{+}_{i} & = & \left\{\phantom{\int}\hspace{-.2in}\{\lambda_{1},\cdots , \lambda_{n_{c}}\} | \ \mathrm{for \ any  \ initial \ subset \ of  \ size} \ \ell_{1} \right. \\
&&\left. \frac{q_{2}}{q_{1}}= \frac{\ell_{2}}{\ell_{1}}, \  \frac{q_{3}}{q_{1}}= \frac{\ell_{3}}{\ell_{1}}, \cdots , \frac{q_{i-1}}{q_{1}}= \frac{\ell_{i-1}}{\ell_{1}},\  \frac{q_{i}}{q_{1}}< \frac{\ell_{i}}{\ell_{1}} \right\} \cap\mathcal{C}^{+}~. \nonumber
\end{eqnarray}
Each of the sets $\upsilon^{+}_{i},$ for $i>1,$ contains $n_{c}$-tuples of partitions satisfying the stability constraint.  The complete list of all partitions in the stable moduli space $\mathcal{M}_{\gamma^{+}}$ is then the union
\begin{equation}
\Upsilon^{+}= \bigcup_{i=2}^{n_{c}}\upsilon^{+}_{i}~. \label{upsilonp}
\end{equation}
A tuple of partitions $\{\lambda_{i}\}$ defines a cell in the stable moduli space if and only if it lies in the set $ \Upsilon^{+}$.

We may now conclude with a complete formula for the dimension of the moduli space $\mathcal{M}_{\gamma}^{+}$. To compute this dimension, we must find those cells of top dimension which maximize \eqref{qposdim1}.  Thus, we have the result
\begin{equation}
\mathrm{dim}\left(\mathcal{M}_{\gamma^{+}}\right)= \max_{\{\lambda_{i}\} \in \Upsilon^{+}}\left[ \phantom{\int}\hspace{-.17in}\Delta_{q>0}(\left\{\lambda_{1}, \cdots \lambda_{n_{c}}\right\})\right]~. \label{qposdim2}
\end{equation}

\subsubsection{Bound States of Dyons with Negative Magnetic Charge}
\label{secqnegreg}

A completely parallel analysis holds for the representations with total negative magnetic charge.  At the $i$-th two node quiver, the representation $K^{i}$ is of the form

\begin{equation}
\begin{xy}  0;<1pt,0pt>:<0pt,-1pt>::
(0,0)*+{\mathbb{C}^{e_{i}}} ="1",
(60,0)*+{\mathbb{C}^{e_{i}-|q_{i}|}} ="2",
(30,9)*+{B_{i}},
(30,-9)*+{A_{i}},
"1", {\ar"2" <2.5pt>},
"1", {\ar"2" <-2.5pt>},
\end{xy}~.
\label{splitrep4}
\end{equation}

The direct sum decomposition of the above is again determined by a partition of $e_{i}$ into $|q_{i}|$ parts.  A slight difference from the analysis in \S \ref{secqposreg} is that now each part must be strictly positive.  Hence again letting $\lambda_{i}$ denote the multiplicities, we write
\begin{equation}
\sum_{s=1}^{e_{i}} \lambda_{i}(s) =|q_{i}|~, \hspace{.5in} \sum_{s=1}^{e_{i}}s \lambda_{i}(s) =e_{i}~. \label{partdata2}
\end{equation}
The partition $\lambda_{i}$ specifies the direct sum decomposition of $K^{i}$ as
\begin{equation}
K^{i}=\left(\bigoplus_{\alpha=1}^{\lambda_{i}(1)}\mathcal{S}_{1}\right) \oplus \left(\bigoplus_{\alpha=1}^{\lambda_{i}(2)}\mathcal{S}_{2}\right) \oplus \cdots \oplus \left(\bigoplus_{\alpha=1}^{\lambda_{i}(e_{i})}\mathcal{S}_{e_{i}}\right) ~. \label{bigsplit1}
\end{equation}

We now introduce the maps $\Phi_{j}$ and by counting their parameters and subtracting automorphisms, we again deduce the dimension of the cell in moduli space associated to the tuple of partitions $\{\lambda_{i}\}.$ Denoting this dimension by $\Delta_{q<0},$ we have
\begin{equation}
\Delta_{q<0}(\left\{\lambda_{1}, \cdots \lambda_{n_{c}}\right\})= \sum_{i=1}^{n_{c}-1}\sum_{r=0}^{e_{i}}\sum_{s=0}^{r}(r-s)\lambda_{i}(r)\lambda_{i+1}(s) -\sum_{i=1}^{n_{c}}\sum_{s=0}^{e_{i}}\lambda_{i}(s)^{2}+1~, \label{qnegdim1}
\end{equation}

Each tuple of partitions yields a potential cell in the moduli space $\mathcal{M}_{\gamma^{-}}$ of charge $\gamma^{-}$ (the superscript $-$ denotes that this is a state of total negative magnetic charge).  However, to enforce stability partitions which do not permit sufficient binding of the dyons must be removed.  

First, we define a connected locus $\mathcal{C}^{-}$ to be those $n_{c}$-tuples of partitions that have the property that for each part $n$ of the partition of $e_{i+1}$, there exists a part $m$ of the partition of $e_{i}$ with $m>n.$  Every $n_{c}$-tuple of partitions which yields a cell in the stable moduli space $\mathcal{M}_{\gamma^{-}}$ must lie in $\mathcal{C}^{-}.$

Next we introduce subsets $\upsilon^{-}_{i}$ of partitions defined as
\begin{eqnarray}
\upsilon^{-}_{i} & = & \left\{\phantom{\int}\hspace{-.2in}\{\lambda_{1},\cdots , \lambda_{n_{c}}\} | \ \mathrm{for \ any  \ initial \ subset \ of  \ size} \ \ell_{1} \right. \\
&&\left. \frac{|q_{2}|}{|q_{1}|}= \frac{\ell_{2}}{\ell_{1}}, \  \frac{|q_{3}|}{|q_{1}|}= \frac{\ell_{3}}{\ell_{1}}, \cdots , \frac{|q_{i-1}|}{|q_{1}|}= \frac{\ell_{i-1}}{\ell_{1}},\  \frac{|q_{i}|}{|q_{1}|}< \frac{\ell_{i}}{\ell_{1}} \right\} \cap\mathcal{C}^{-}~. \nonumber
\end{eqnarray}

The complete set of all $n_{c}$-tuples of partitions that correspond to cells in the stable moduli space is given by
\begin{equation}
\Upsilon^{-}= \bigcup_{i=2}^{n_{c}}\upsilon^{-}_{i}~. \label{upsilonm}
\end{equation}

Finally, the dimension of the stable moduli space $\mathcal{M}_{\gamma^{-}}$ is then given by extracting the dimension of the top cell
\begin{equation}
\mathrm{dim}\left(\mathcal{M}_{\gamma^{-}}\right)= \max_{\{\lambda_{i}\}\in \Upsilon^{-}}\left[ \phantom{\int}\hspace{-.17in}\Delta_{q<0}(\left\{\lambda_{1}, \cdots \lambda_{n_{c}}\right\})\right]~.\label{qnegdim2}
\end{equation}

\subsubsection{States with Small Magnetic Charges}
\label{secsmallmag}

The formulas \eqref{qposdim2}-\eqref{qnegdim2} are the final results for the exact dimension of moduli space.  In general, these dimensions are complicated arithmetic functions of the charges $\{e_{i},q_{i}\}$ describing the bound state particle.  Physically speaking, this means that the multi-centered configuration of dyons which maximizes the angular momentum depends in detail on the total charge in question.

To understand the results, it is instructive to evaluate the dimension formulas in simple examples involving small charges.

\paragraph{Total positive magnetic charges: $(q_{1}, q_{2}, \cdots, q_{n_{c}-1},q_{n_{c}})=(1,1,\cdots,1,q_{n_{c}})$}\mbox{}

Consider the special of positive magnetic charge bound states where the $q_{i}$ are restricted as
\begin{equation}
(q_{1}, q_{2}, \cdots, q_{n_{c}-1},q_{n_{c}})=(1,1,\cdots,1,q_{n_{c}})~.
\end{equation}
The partitions (not the multiplicities $\lambda_{i}$) are then simply
\begin{equation}
\{e_{1}\}~, \hspace{.2in}\{e_{2}\}~, \hspace{.2in}\cdots \hspace{.2in} \{e_{n_{c}-1}\}~, \hspace{.2in}\{x_{1}, x_{2}, \cdots, x_{q_{n_{c}}}\}~,
\end{equation}
where the sum of the $x_{i}$ is $e_{n_{c}}$.  

The stability constraints, that the collection of partitions lie in $\Upsilon^{+}$ of \eqref{upsilonp}, are simple to evaluate.  They imply merely that the representation does not split into a direct sum.  In other words, each element of $(i+1)$-th partition, must lie in the image of all elements of the $i$-th partition.  Whence, the partition of $e_{n_{c}}$ is constrained to satisfy
\begin{equation}
e_{n_{c}-1} < x_{i}~, \hspace{.5in} \forall i~, \label{stabxi}
\end{equation}
and hence stable bound states with these charges can exist if and only if
\begin{equation}
e_{1}<e_{2}< \cdots < e_{n_{c}-1}~, \hspace{.5in} \mathrm{and} \hspace{.5in} (1+q_{n_{c}})e_{n_{c}-1} \leq e_{n_{c}}~. \label{cineq1}
\end{equation}
If further we assume that
\begin{equation}
\frac{q_{n_{c}}(q_{n_{c}}+1)}{2} +q_{n_{c}}e_{n_{c}-1}\leq e_{c}~, \label{cineq2}
\end{equation}
then the elements $x_{i}$ of the partition of $e_{c}$ may taken to be distinct.  In that case, any partitions which satisfy \eqref{stabxi} with the $x_{i}$ all distinct maximize the cell dimension function $\Delta_{q>0}$ of equation  \eqref{qposdim1}.  We deduce that 
\begin{eqnarray}
\mathrm{dim}\left(\mathcal{M}_{\gamma^{+}}\right) & = & \sum_{i=1}^{n_{c}-2} (e_{i+1}-e_{i})+\sum_{j=1}^{q_{n_{c}}}(x_{j}-e_{n_{c}-1})-(n_{c}-1)-q_{n_{c}}+1~, \nonumber\\
& =& e_{n_{c}}-(q_{n_{c}}-1)e_{n_{c}-1}-e_{1}-q_{n_{c}}-(n_{c}-2)~,\\ 
& \geq & \frac{q_{n_{c}}(q_{n_{c}}-1)}{2}+\left[e_{n_{c}-1}-(e_{1}+n_{c}-2)\right]~, \nonumber
\end{eqnarray}
where in the last step, we used the charge inequalities \eqref{cineq1}-\eqref{cineq2} to illustrate that $\mathrm{dim}\left(\mathcal{M}_{\gamma^{+}}\right)$ is non-negative.

\begin{figure}[h!]
  \centering
  \includegraphics[width=0.4\textwidth]{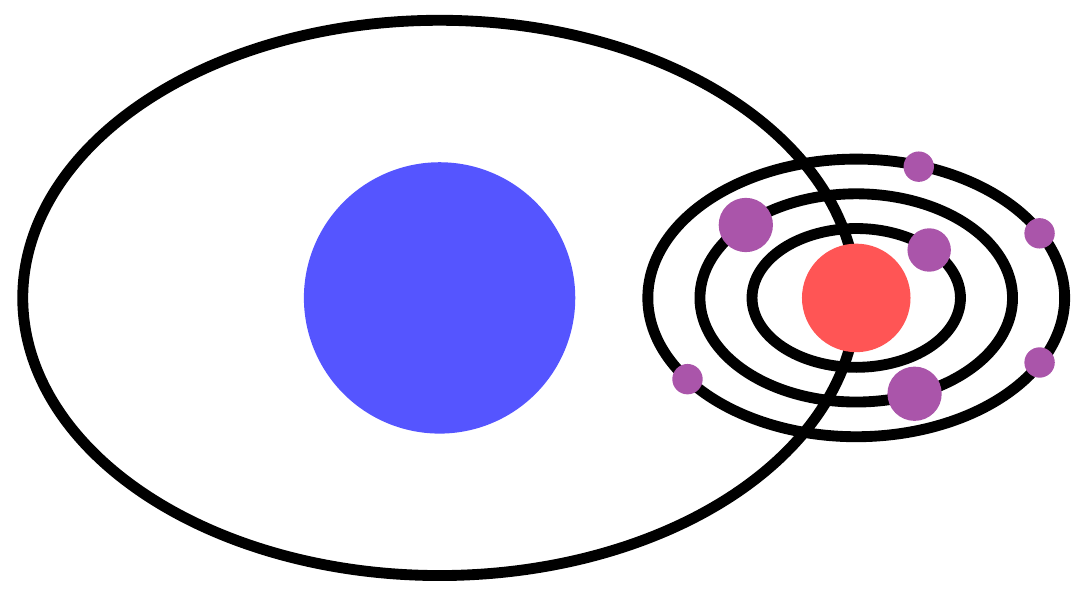}
  \caption{A cartoon of the bound states with $n_{c}=3$ and $(q_{1},q_{2}, q_{3})=(1,1,7)$.  The central blue object is the heavy dyon carrying a single unit of $q_{1}$ charge.  The red object is the dyon with $q_{2}=1$.  This object is parametrically lighter than the central dyon.  Finally, we have various purple dyons carrying charge $q_{3}=1$.  These are parametrically lighter than red dyon that they orbit.  }
  \label{fig:egg1}
\end{figure}

Since the angular momentum is controlled by the dimension of moduli space as in \eqref{jdim}, we now readily determine that the state of maximal angular momentum with these electric and magnetic charges has
\begin{equation}
J=\frac{1}{2}\left(e_{n_{c}}-(q_{n_{c}}-1)e_{n_{c}-1}-e_{1}-q_{n_{c}}-(n_{c}-2)\right)~.
\end{equation}
In particular, by increasing the electric charges, we make states whose angular momentum is arbitrarily large.  Meanwhile, we may also estimate the mass of these states.  Indeed, in the approximation of interest the dominant contribution to the mass comes from the rest energy of the dyon with $q_{1}=1$.  Thus, these states have mass $m\sim |\zeta_{1}|$ plus small corrections.  A cartoon of these states is illustrated in Figure \ref{fig:egg1}.

One may similarly obtain exact formulas for the angular momentum of BPS states where the magnetic charges satisfy the property that if $q_{i}>1,$ then $q_{i-1}=q_{i+1}=1$.

\subsubsection{Semiclassical Estimate of the Regge Slope}
\label{secsemi}

For larger magnetic charges it is cumbersome to evaluate \eqref{qposdim1}-\eqref{qnegdim1} to obtain exact results for the dimension of moduli space.  Nevertheless, we may obtain a simple estimate which enables us to demonstrate Regge growth in the angular momentum of the BPS states at weak coupling.

We examine again the cell dimension formulas \eqref{qposdim1}-\eqref{qnegdim1}:
\begin{eqnarray}
\Delta_{q>0}(\left\{\lambda_{1}, \cdots \lambda_{n_{c}}\right\}) & = & \sum_{i=1}^{n_{c}-1}\sum_{r=0}^{e_{i+1}}\sum_{s=0}^{r}(r-s)\lambda_{i+1}(r)\lambda_{i}(s) -\sum_{i=1}^{n_{c}}\sum_{s=0}^{e_{i}}\lambda_{i}(s)^{2}+1~,\nonumber \\
\Delta_{q<0}(\left\{\lambda_{1}, \cdots \lambda_{n_{c}}\right\}) &= & \sum_{i=1}^{n_{c}-1}\sum_{r=0}^{e_{i}}\sum_{s=0}^{r}(r-s)\lambda_{i}(r)\lambda_{i+1}(s) -\sum_{i=1}^{n_{c}}\sum_{s=0}^{e_{i}}\lambda_{i}(s)^{2}+1~.\label{dimlast}
\end{eqnarray}
If a tuple of partitions $\left\{\lambda_{1}, \cdots \lambda_{n_{c}}\right\}$ lies in $\Upsilon^{+}$ or $\Upsilon^{-},$ then as described in \S \ref{secqposreg}-\ref{secqnegreg}, it contributes a cell to the stable moduli space whose dimensions are computed by the above formulas.  

To obtain an estimate on the dimension of moduli space, we must first determine a simple class of $n_{c}$-tuples of partitions which always lie in the stable moduli space. From the definitions of the sets $\Upsilon^{\pm},$ it is straightforward to deduce the following. 
\begin{itemize}
\item For representations with total positive magnetic charge, the $n_{c}$-tuples of partitions where each element of the $(i+1)$-th partition is strictly greater than each element of the $i$-th partition, always lie in the stable set $\Upsilon^{+}$.
\item For representations with total negative magnetic charge, the $n_{c}$-tuples of partitions where each element of the $(i+1)$-th partition is strictly less than each element of the $i$-th partition, always lie in the stable set $\Upsilon^{-}$.
\end{itemize}
For reasons which shall become clear momentarily, we refer to a tuple of partitions satisfying either of the above properties as \emph{regular}.

It is significant that the condition of being regular behaves well under scaling.  If $\{\lambda_{i}\}$ is a tuple of partitions associated to charge $\{e_{i},q_{i}\}$, and the tuple $\{\lambda_{i}\}$ is regular, then there exist tuples of partitions  associated to charges $\{\Lambda e_{i}, \Lambda q_{i}\},$  for $\Lambda >1$ which are also regular.  

Moreover it is also easy to determine sufficient conditions on charges $\{e_{i},q_{i}\}$ for regular tuples of partitions to exist.  Denote by $\lceil x \rceil,$ and $\lfloor x \rfloor,$ the ceiling and floor of a real number $x$.  Then we have:
\begin{itemize}
\item For representations with all positive magnetic charges, a sufficient condition for regular tuples of partitions to exist is
\begin{equation}
\lceil e_{i}/q_{i}\rceil < \lfloor e_{i+1}/q_{i+1}\rfloor~, \hspace{.5in} i=1,\cdots, n_{c}-1~.
\end{equation}
\item For representations with all negative magnetic charges, a sufficient condition for regular tuples of partitions to exist is
\begin{equation}
\lfloor e_{i}/|q_{i}|\rfloor > \lceil e_{i+1}/|q_{i+1}|\rceil~, \hspace{.5in} i=1,\cdots, n_{c}-1~.
\end{equation}
\end{itemize}

The robust stability properties of regular tuples of partitions enable us to use them to investigate the stable moduli space $\mathcal{M}_{\gamma^{\pm}}$ for large charges.

In addition to the stability properties outlined above, the significance of the regularity property for a tuple of partitions, is that the dimension formulas \eqref{dimlast} simplify dramatically.  Indeed on any regular tuple of partitions we have 
\begin{eqnarray}
\Delta_{q \overset{>}{<}0}(\left\{\lambda_{1}, \cdots \lambda_{n_{c}}\right\}) & = & \sum_{i=1}^{n_{c}-1}(e_{i+1}q_{i}-e_{i}q_{i+1}) -\sum_{i=1}^{n_{c}}\sum_{s=0}^{e_{i}}\lambda_{i}(s)^{2}+1~.\label{dimlast1}
\end{eqnarray}
In particular, the only difference between positive and negative magnetic charges is whether $q_{i}$ are positive or negative in the above.

The simplified form of $\Delta$ given in \eqref{dimlast1} has an elementary physical interpretation.   The quadratic form on the charges appearing in \eqref{dimlast1} is the sum of the symplectic products of each of the $K^{i},$ the dyons in each $SU(2)$ subgroup.  Indeed, letting $\gamma_{K^{i}}$ denote the charge vector for $K^{i},$ and using the pairing \eqref{cartanpairing} we have
\begin{equation}
\sum_{i=1}^{n_{c}-1}(e_{i+1}q_{i}-e_{i}q_{i+1})= \langle\gamma_{K^{1}} , \gamma_{K^{2}}\rangle +\langle \gamma_{K^{2}},  \gamma_{K^{3}}\rangle+\cdots +\langle\gamma_{K^{n_{c}-1}}, \gamma_{K^{n_{c}}}\rangle ~.
\end{equation}
This is the naive semiclassical contribution to the angular momentum discussed in \S \ref{secintuit}.  Each group of dyons $K^{i}$ binds to its neighbor $K^{i+1}$ and in so doing sources an angular momentum proportional to the Dirac pairing $ \langle\gamma_{K^{i}} , \gamma_{K^{i+1}}\rangle$.   

Meanwhile, the negative contribution in \eqref{dimlast1} associated to multiplicity in the partitions can be understood as a quantum correction to the semiclassical intuition, which arises due to the quantum statistics of the identical constituents making up the $K^{i}$.

In summary, regular tuples of partitions yield cells in the quiver moduli spaces $\mathcal{M}_{\gamma^{\pm}}$ with readily understandable dimensions.  What our detailed analysis of stability conditions has yielded is thus sufficient conditions for such bound states to exist.

To estimate the dimension of moduli space, we must now determine the importance of the multiplicity subtractions in \eqref{dimlast1}.  Consider a pair $(e_{i}, q_{i})$.  In order for there to admit a partition of $e_{i}$ into $q_{i}$ parts without any repeated entry, one must have (for large $q_{i}$)
\begin{equation} 
e_{i} > \frac{1}{2}q_{i}^{2}~.
\end{equation}
Significantly, the above is inhomogeneous under rescaling $(e_{i},q_{i})\rightarrow (\Lambda e_{i}, \Lambda q_{i}).$  It follows that, if we examine any direction in the charge lattice $\hat{\gamma},$ then sufficiently far our along the ray $\Lambda \hat{\gamma}$, the multiplicity term in \eqref{dimlast1} yields an important correction to the dimension of the associated cell.

We can easily take these considerations into account and produce a lower bound on the dimension of moduli space.  The largest subtraction that can occur as a result of multiplicity is when all terms in the partition are equal in which case the multiplicity subtractions evaluate to simply $\sum_{i}q_{i}^{2}.$  Combining this with the fact that the semiclassical partitions provide a lower bound on possible cell dimensions, we obtain 
\begin{equation}
\mathrm{dim}(\mathcal{M}_{\gamma})\geq \sum_{i=1}^{n_{c}-1}(e_{i+1}q_{i}-e_{i}q_{i+1})-\sum_{i=1}^{n_{c}}q_{i}^{2}+1~. \label{dimest}
\end{equation}

To summarize our conclusions.
\begin{itemize}
\item If $q_{i}>0$ for all $i,$ and $\lceil e_{i}/q_{i}\rceil < \lfloor e_{i+1}/q_{i+1}\rfloor$ for $i=1,\cdots, n_{c}-1,$ then, in the weak coupling region of moduli space, there exist stable BPS particles with the given charges $\{e_{i},q_{i}\}$.  
 \item If $q_{i}<0$ for all $i,$ and  $\lfloor e_{i}/|q_{i}|\rfloor > \lceil e_{i+1}/|q_{i+1}|\rceil$ for $i=1,\cdots, n_{c}-1,$ then, in the weak coupling region of moduli space, there exist stable BPS particles with the given charges $\{e_{i},q_{i}\}$.  
 \item In either of the two cases described above, we have a lower bound on the angular momentum
\begin{equation}
J\geq \frac{1}{2}\left(\sum_{i=1}^{n_{c}-1}(e_{i+1}q_{i}-e_{i}q_{i+1})-\sum_{i=1}^{n_{c}}q_{i}^{2}\right)+\frac{1}{2}~.
\end{equation}
\end{itemize}

Since the angular momentum in our estimate scales quadratically with charges, we immediately determine that the BPS spectrum shows Regge behavior.  Indeed expressing the angular momentum as
\begin{equation}
J\geq  \frac{\left(\sum_{i=1}^{n_{c}-1}(e_{i+1}q_{i}-e_{i}q_{i+1})-\sum_{i=1}^{n_{c}}q_{i}^{2}\right)}{2|q_{1}\zeta_{1}+q_{2}\zeta_{2}+\cdots+q_{n_{c}}\zeta_{n_{c}}|^{2}} m^{2}+\frac{1}{2}~,
\end{equation}
We read an estimate for the slope as 
\begin{equation}
\alpha'\geq\left[\frac{\left(\sum_{i=1}^{n_{c}-1}(e_{i+1}q_{i}-e_{i}q_{i+1})-\sum_{i=1}^{n_{c}}q_{i}^{2}\right)}{2|q_{1}\zeta_{1}+q_{2}\zeta_{2}+\cdots+q_{n_{c}}\zeta_{n_{c}}|^{2}} \right]~.
\end{equation}
In particular, the the above bound for the slope, depends only on the direction $\hat{\gamma}$ in the charge lattice, as anticipated.

We may perform one simple consistency check on our result.  The bare parameters in the Lagrangian defining the $SU(n_{c}+1)$ SYM theory, may be packaged into the complex strong coupling scale $\Lambda.$  A redefinition $\Lambda \rightarrow e^{2\pi i}\Lambda$ leaves the theory invariant, but causes a monodromy in the definitions of charges
\begin{equation}
e_{i}\rightarrow e_{i}+q_{i}~, \hspace{.5in} q_{i}\rightarrow q_{i}~.
\end{equation}
Thus, the weak-coupling spectrum, and in particular our angular momentum bound, must be invariant under this transformation.  Happily, it is.

\section*{Acknowledgements} 
I would like to thank S.H. Shao for comments on a draft.  My work is support by a Junior Fellowship at the Harvard Society of Fellows.

\appendix
\addtocontents{toc}{\protect\setcounter{tocdepth}{1}}

\section{Gluing Constraints}
\label{glueme}

In this section we solve the superpotential constraints, that appearing by varying \eqref{ncnfw} with respect to $\Phi_{i}$ or $\Psi_{j}$.  These results are instrumental in obtaining the conclusions of \S \ref{supsolve}.  Throughout, we make frequent use of \S \ref{secsu2reps} which provides a detailed description of the maps $A_{i}$ and $B_{i}$.  We refer to the remaining map ($\Phi$ or $\Psi$) as $\phi$, the gluing map.  Our aim is to derive the number of complex parameters in gluing maps between the various possible $SU(2)$ representations.

\subsection{Gluing $(n+1,n)$ to $(\ell+1, \ell)$}
Consider the representation illustrated below.
\begin{equation}
\begin{xy}  0;<1pt,0pt>:<0pt,-1pt>::
(0,0)*+{\mathbb{C}^{n+1}} ="1",
(60,0)*+{\mathbb{C}^{n}} ="2",
(30,10)*+{B_{1}},
(30,-10)*+{A_{1}},
(90,-10)*+{\phi},
(120,0)*+{\mathbb{C}^{\ell+1}} ="3",
(180,0)*+{\mathbb{C}^{\ell}} ="4",
(150,10)*+{B_{2}},
(150,-10)*+{A_{2}},
"1", {\ar"2" <2.5pt>},
"1", {\ar"2" <-2.5pt>},
"2", {\ar"3" <0pt>},
"3", {\ar"4" <2.5pt>},
"3", {\ar"4" <-2.5pt>},
\end{xy}
\label{Partial1}
\end{equation}
At each to the two Kronecker subquivers we may introduce preferred bases 
\begin{equation}
\mathrm{span}\{v_{1}, v_{2}, \cdots, v_{n+1}\}= \mathbb{C}^{n}~, \hspace{.5in}\mathrm{span}\{w_{1}, w_{2}, \cdots, w_{n}\}= \mathbb{C}^{n}~,
\end{equation}
\begin{equation}
\mathrm{span}\{x_{1}, x_{2}, \cdots, x_{\ell+1}\}= \mathbb{C}^{m}~, \hspace{.5in}\mathrm{span}\{y_{1}, y_{2}, \cdots, y_{\ell}\}= \mathbb{C}^{\ell}~. \nonumber
\end{equation}

The linear maps may be put in simple form as 
\begin{equation}
A_{1}(v_{i})= \begin{cases} w_{i} & i\leq n \\ 0 & i=n+1\end{cases}~, \hspace{.5in}B_{1}(v_{i})= \begin{cases} 0 & i=1 \\ w_{i-1} & i>1\end{cases}~.
\end{equation}
And
\begin{equation}
A_{2}(x_{i})= \begin{cases} y_{i} & i\leq \ell \\ 0 & i=\ell+1\end{cases}~, \hspace{.5in}B_{2}(x_{i})= \begin{cases} 0 & i=1 \\ y_{i-1} & i>1\end{cases}~.
\end{equation}

Now consider the behavior of the map $\phi.$  Introduce matrix notation as
\begin{equation}
\phi(w_{j})=\sum_{s=1}^{\ell+1}\phi_{s,j}x_{s}~, \hspace{.5in}j \leq n~.
\end{equation}
The complex parameters $\phi_{s,j}$ must be chosen to solve the constraint
\begin{equation}
A_{2}\circ \phi \circ A_{1}=B_{2} \circ \phi \circ B_{1}~.
\end{equation}
Evaluated on the vector $v_{i}$ this yields
\begin{equation}
\sum_{s=1}^{\ell}\phi_{s,i}y_{s}=\sum_{s=2}^{\ell+1}\phi_{s,i-1}y_{s-1}~, \hspace{.5in} 1<i\leq n,~
\end{equation}
as well as
\begin{equation}
0=\sum_{s=2}^{\ell+1}\phi_{s,n}y_{s-1}~, \hspace{.5in}0=\sum_{s=1}^{\ell}\phi_{s,1}y_{s}~.
\end{equation}
Matching coefficients, we see that this amounts to
\begin{equation}
\phi_{s+1,n}=0~, \hspace{.5in}\phi_{s,1}=0~, \hspace{.5in}\phi_{s,i}=\phi_{s+1,i-1}~,\hspace{.5in}  \forall s \leq \ell,  \ \mathrm{and}  \ \ 2\leq i\leq n.
\end{equation}
We readily solve these equations to deduce the number of parameters in the map $\phi$.  For instance if $n=5,$ and $\ell=3$ we have
\begin{equation}
\phi=\left(\begin{array}{ccccc}0 & 0 & \alpha & \beta & \gamma \\0 & \alpha & \beta & \gamma & 0 \\ \alpha  & \beta & \gamma & 0 & 0\end{array}\right)~,
\end{equation}
Where $\alpha, \beta, \gamma$ are arbitrary complex parameters.  In general we conclude the following.

\begin{prop} \label{Count1} The number of parameters for gluing $(n+1,n)$ to $(\ell+1, \ell)$ is $n-\ell$.  In particular, the gluing map must vanish if $n\leq \ell$.  Moreover, if the map $\phi$ is non-zero then it is surjective.
\end{prop}

\subsection{Gluing $(n+1,n)$ to $(\ell, \ell)$}
Next consider representations of the form illustrated below.
\begin{equation}
\begin{xy}  0;<1pt,0pt>:<0pt,-1pt>::
(0,0)*+{\mathbb{C}^{n+1}} ="1",
(60,0)*+{\mathbb{C}^{n}} ="2",
(30,10)*+{B_{1}},
(30,-10)*+{A_{1}},
(90,-10)*+{\phi},
(120,0)*+{\mathbb{C}^{\ell}} ="3",
(180,0)*+{\mathbb{C}^{\ell}} ="4",
(150,10)*+{B_{2}},
(150,-10)*+{A_{2}},
"1", {\ar"2" <2.5pt>},
"1", {\ar"2" <-2.5pt>},
"2", {\ar"3" <0pt>},
"3", {\ar"4" <2.5pt>},
"3", {\ar"4" <-2.5pt>},
\end{xy}
\label{Partial2}
\end{equation}
At each to the two Kronecker subquivers we may introduce preferred bases 
\begin{equation}
\mathrm{span}\{v_{1}, v_{2}, \cdots, v_{n+1}\}= \mathbb{C}^{n}~, \hspace{.5in}\mathrm{span}\{w_{1}, w_{2}, \cdots, w_{n}\}= \mathbb{C}^{n}~,
\end{equation}
\begin{equation}
\mathrm{span}\{x_{1}, x_{2}, \cdots, x_{\ell}\}= \mathbb{C}^{m}~, \hspace{.5in}\mathrm{span}\{y_{1}, y_{2}, \cdots, y_{\ell}\}= \mathbb{C}^{\ell}~. \nonumber
\end{equation}
The maps $A_{1}$ and $B_{1}$ take the form
\begin{equation}
A_{1}(v_{i})= \begin{cases} w_{i} & i\leq n \\ 0 & i=n+1\end{cases}~, \hspace{.5in}B_{1}(v_{i})= \begin{cases} 0 & i=1 \\ w_{i-1} & i>1\end{cases}~,
\end{equation}
while $A_{2}$ and $B_{2}$ are
\begin{equation}
A_{2}(x_{i})= \begin{cases} \lambda y_{i} & i=1 \\ \lambda y_{i}+y_{i-1} &i >1\end{cases}~, \hspace{.5in}B_{1}(x_{i})=y_{i}~.
\end{equation}

We now consider $\phi.$  Introduce matrix notation as
\begin{equation}
\phi(w_{j})=\sum_{s=1}^{\ell}\phi_{s,j}x_{s}~, \hspace{.5in}j \leq n~.
\end{equation}
The parameters $\phi_{s,j}$ must be chosen to solve
\begin{equation}
A_{2}\circ \phi \circ A_{1}=B_{2} \circ \phi \circ B_{1}~.
\end{equation}
Evaluated on our preferred bases this becomes
\begin{equation}
\lambda \phi_{s,i}+\phi_{s+1,i}=\phi_{s,i-1}~, \hspace{.5in} 1<i \leq n  \ \mathrm{and}  \ \ 1\leq s <\ell~.
\end{equation}
As well as
\begin{equation}
\phi_{s,n}=0~,  \hspace{.25in} 1\leq s \leq \ell~, \hspace{.5in}  \lambda \phi_{\ell,i}= \phi_{\ell,i-1}~,\hspace{.25in} 1<i \leq n.
\end{equation}
These constraints may only be satisfied if all $\phi_{s,i}$ vanish.  Thus we conclude
\begin{prop} \label{Count2} The representation $(n+1,n)$ cannot be glued to the representation $(\ell, \ell).$ 
\end{prop}
\subsection{Gluing $(n+1,n)$ to $(\ell, \ell+1)$}
Next we examine representations of the form
\begin{equation}
\begin{xy}  0;<1pt,0pt>:<0pt,-1pt>::
(0,0)*+{\mathbb{C}^{n+1}} ="1",
(60,0)*+{\mathbb{C}^{n}} ="2",
(30,10)*+{B_{1}},
(30,-10)*+{A_{1}},
(90,-10)*+{\phi},
(120,0)*+{\mathbb{C}^{\ell}} ="3",
(180,0)*+{\mathbb{C}^{\ell+1}} ="4",
(150,10)*+{B_{2}},
(150,-10)*+{A_{2}},
"1", {\ar"2" <2.5pt>},
"1", {\ar"2" <-2.5pt>},
"2", {\ar"3" <0pt>},
"3", {\ar"4" <2.5pt>},
"3", {\ar"4" <-2.5pt>},
\end{xy}
\label{Partial3}
\end{equation}

As before we introduce bases
\begin{equation}
\mathrm{span}\{v_{1}, v_{2}, \cdots, v_{n+1}\}= \mathbb{C}^{n+1}~, \hspace{.5in}\mathrm{span}\{w_{1}, w_{2}, \cdots, w_{n}\}= \mathbb{C}^{n}~,
\end{equation}
\begin{equation}
\mathrm{span}\{x_{1}, x_{2}, \cdots, x_{\ell}\}= \mathbb{C}^{\ell}~, \hspace{.5in}\mathrm{span}\{y_{1}, y_{2}, \cdots, y_{\ell+1}\}= \mathbb{C}^{\ell+1}~. \nonumber
\end{equation}
The maps $A_{1}$ and $B_{1}$ take the form
\begin{equation}
A_{1}(v_{i})= \begin{cases} w_{i} & i\leq n \\ 0 & i=n+1\end{cases}~, \hspace{.5in}B_{1}(v_{i})= \begin{cases} 0 & i=1 \\ w_{i-1} & i>1\end{cases}~,
\end{equation}
while $A_{2}$ and $B_{2}$ are given by
\begin{equation}
A_{2}(x_{i})=y_{i}, \hspace{.5in}B_{2}(x_{i})=y_{i+1}~.
\end{equation}

The gluing map $\phi$ is defined by its matrix elements 
\begin{equation}
\phi(w_{j})=\sum_{s=1}^{\ell}\phi_{s,j}x_{s}~, \hspace{.5in}j \leq n~.
\end{equation}
It is constrained by requiring
\begin{equation}
A_{2}\circ \phi \circ A_{1}=B_{2} \circ \phi \circ B_{1}~.
\end{equation}
In terms of the matrix elements of $\phi,$ this amounts to
\begin{equation}
\phi_{s,n}=0~, \hspace{.5in} \phi_{s,1}=0~, \hspace{.5in}\forall s~.
\end{equation}
As well as
\begin{equation}
\phi_{1,i}=0~, \hspace{.5in}\phi_{\ell,i-1}=0~, \hspace{.5in} \phi_{s,i}=\phi_{s-1,i-1}~, \hspace{.5in}1<i<n+1~, \mathrm{and} \ 1<s<\ell+1~. 
\end{equation}
The only solution to these constraints has the entire map $\phi$ vanishing.  Thus we conclude
\begin{prop} \label{Count3} The representation $(n+1,n)$ cannot be glued to the representation $(\ell, \ell+1).$ 
\end{prop}

\subsection{Gluing $(n,n)$ to $(\ell, \ell+1)$}
Next consider the representation illustrated below
\begin{equation}
\begin{xy}  0;<1pt,0pt>:<0pt,-1pt>::
(0,0)*+{\mathbb{C}^{n}} ="1",
(60,0)*+{\mathbb{C}^{n}} ="2",
(30,10)*+{B_{1}},
(30,-10)*+{A_{1}},
(90,-10)*+{\phi},
(120,0)*+{\mathbb{C}^{\ell}} ="3",
(180,0)*+{\mathbb{C}^{\ell+1}} ="4",
(150,10)*+{B_{2}},
(150,-10)*+{A_{2}},
"1", {\ar"2" <2.5pt>},
"1", {\ar"2" <-2.5pt>},
"2", {\ar"3" <0pt>},
"3", {\ar"4" <2.5pt>},
"3", {\ar"4" <-2.5pt>},
\end{xy}
\label{Partial4}
\end{equation}

We introduce bases as
\begin{equation}
\mathrm{span}\{v_{1}, v_{2}, \cdots, v_{n}\}= \mathbb{C}^{n}~, \hspace{.5in}\mathrm{span}\{w_{1}, w_{2}, \cdots, w_{n}\}= \mathbb{C}^{n}~,
\end{equation}
\begin{equation}
\mathrm{span}\{x_{1}, x_{2}, \cdots, x_{\ell}\}= \mathbb{C}^{\ell}~, \hspace{.5in}\mathrm{span}\{y_{1}, y_{2}, \cdots, y_{\ell+1}\}= \mathbb{C}^{\ell+1}~. \nonumber
\end{equation}
The maps $A_{1}$ and $B_{1}$ take the form
\begin{equation}
A_{1}(v_{i})= \begin{cases} \lambda w_{i} & i=1 \\ \lambda w_{i}+w_{i-1} &i >1\end{cases}~, \hspace{.5in}B_{1}(v_{i})=w_{i}~.
\end{equation}
while $A_{2}$ and $B_{2}$ are given by
\begin{equation}
A_{2}(x_{i})=y_{i}~, \hspace{.5in}B_{2}(x_{i})=y_{i+1}~.
\end{equation}

The map $\phi$ is defined by its matrix elements
\begin{equation}
\phi(w_{j})=\sum_{s=1}^{\ell}\phi_{s,j}x_{s}~, \hspace{.5in} j\leq n~.
\end{equation}

We must choose $\phi_{s,j}$ to satisfy the constraint
\begin{equation}
A_{2}\circ \phi \circ A_{1}= B_{2}\circ \phi \circ B_{1}~.
\end{equation}
Evaluating on our bases above this constraint amounts to
\begin{equation}
\phi_{s,1}=0~, \hspace{.5in} \forall s~, \hspace{.5in}\phi_{\ell,i}=0~, \hspace{.5in}\forall i~.
\end{equation}
And
\begin{equation}
\phi_{s-1,i}=\lambda \phi_{s,i}+\phi_{s,i-1}~, \hspace{.5in} 2\leq i \leq n~,  \ \mathrm{and}  \ \ 2\leq s\leq \ell~.
\end{equation}

The only solution to these constraints has all matrix elements $\phi_{s,i}$ vanishing.  Thus we conclude:
\begin{prop} \label{Count4} The representation $(n,n)$ cannot be glued to the representation $(\ell, \ell+1).$ 
\end{prop}

\subsection{Gluing $(n,n+1)$ to $(\ell,\ell+1)$}

Consider the representation illustrated below.
\begin{equation}
\begin{xy}  0;<1pt,0pt>:<0pt,-1pt>::
(0,0)*+{\mathbb{C}^{n}} ="1",
(60,0)*+{\mathbb{C}^{n+1}} ="2",
(30,10)*+{B_{1}},
(30,-10)*+{A_{1}},
(90,-10)*+{\phi},
(120,0)*+{\mathbb{C}^{\ell}} ="3",
(180,0)*+{\mathbb{C}^{\ell+1}} ="4",
(150,10)*+{B_{2}},
(150,-10)*+{A_{2}},
"1", {\ar"2" <2.5pt>},
"1", {\ar"2" <-2.5pt>},
"2", {\ar"3" <0pt>},
"3", {\ar"4" <2.5pt>},
"3", {\ar"4" <-2.5pt>},
\end{xy}
\label{Partial5}
\end{equation}
At each to the two Kronecker subquivers we may introduce preferred bases 
\begin{equation}
\mathrm{span}\{v_{1}, v_{2}, \cdots, v_{n}\}= \mathbb{C}^{n}, \hspace{.5in}\mathrm{span}\{w_{1}, w_{2}, \cdots, w_{n+1}\}= \mathbb{C}^{n+1},
\end{equation}
\begin{equation}
\mathrm{span}\{x_{1}, x_{2}, \cdots, x_{\ell}\}= \mathbb{C}^{\ell}, \hspace{.5in}\mathrm{span}\{y_{1}, y_{2}, \cdots, y_{\ell+1}\}= \mathbb{C}^{\ell+1}. \nonumber
\end{equation}
The linear maps are again simple
\begin{equation}
A_{1}(v_{i})=w_{i}, \hspace{.5in}B_{1}(v_{i})=w_{i+1}, \hspace{.5in}A_{2}(x_{i})=y_{i}, \hspace{.5in}B_{2}(x_{i})=y_{i+1}.
\end{equation}
Now consider the behavior of the map $\phi.$  Introduce matrix notation as
\begin{equation}
\phi(w_{j})=\sum_{s=1}^{\ell}\phi_{s,j}x_{s}, \hspace{.5in}j \leq n+1.
\end{equation}
The complex parameters $\phi_{s,j}$ must be chosen to solve the constraint
\begin{equation}
A_{2}\circ \phi \circ A_{1}=B_{2} \circ \phi \circ B_{1}.
\end{equation}
Evaluated on the vector $v_{i}$ this yields
\begin{equation}
\sum_{s=1}^{\ell}\phi_{s,i}y_{s}=\sum_{s=1}^{\ell}\phi_{s,i+1}y_{s+1}, \hspace{.5in} \forall i \leq n.
\end{equation}
Matching coefficients of the basis vectors we deduce that the above amounts to
\begin{equation}
\phi_{1,i}=0, \hspace{.5in} \phi_{\ell,i+1}=0, \hspace{.5in}\phi_{s,i}=\phi_{s-1,i+1}~, \hspace{.5in} \forall i \leq n,  \ \mathrm{and}  \ \ 2\leq s\leq \ell.
\end{equation}
We readily solve these constraints, and deduce the number of parameters in the map $\phi$.  For instance with $n=2$ and $\ell=5$ the matrix for $\phi$ takes the form
\begin{equation}
\phi=\left(\begin{array}{ccc}0 & 0 & \alpha \\ 0 & \alpha & \beta \\ \alpha & \beta & \gamma \\ \beta & \gamma & 0 \\ \gamma & 0 & 0 \end{array}\right).
\end{equation}
In general we conclude the following.
\begin{prop} \label{Count5} The number of parameters for gluing $(n,n+1)$ to $(\ell,\ell+1)$ is $\ell-n$.  In particular, the gluing map must vanish if $\ell\leq n$.  If the map $\phi$ is non-zero then it is injective.
\end{prop}

\bibliography{Regge}
\bibliographystyle{utphys}
\end{document}